\documentclass[a4paper,fleqn,usenatbib]{mnras}
\usepackage{amsmath}
\usepackage{mathptmx}
\usepackage{txfonts}
\usepackage[T1]{fontenc}
\usepackage{ae,aecompl}
\usepackage{graphicx}	
\usepackage{amssymb}	
\usepackage[normalem]{ulem}
\usepackage{longtable}
\usepackage{ae,aecompl}
\newcommand{\e}{\end{equation}}
\newcommand{\bear}{\begin{eqnarray}}
\newcommand{\ear}{\end{eqnarray}}
\newcommand{\hmpc}{{\, h^{-1}\, {\rm Mpc}}}
\newcommand{\neff}{N_{\rm effective}}

\def\aj{AJ}
\def\apj{ApJ}
\def\apjs{ApJS}
\def\jcap{JCAP}
\def\mnras{MNRAS}
\def\prl{Physical Review Letters}
\def\aap{A\&A}
\def\prd{Physical Review D}
\def\nat{Nature}
\def\apjs{ApJS}
\def\apjl{ApJ Letters}

\title[Testing isotropy in the SDSS]{Testing isotropy in the Universe
  using photometric and spectroscopic data from the SDSS} \author
      [Sarkar, Pandey \& Khatri] {Suman
        Sarkar$^{1}$\thanks{suman2reach@gmail.com}, Biswajit
        Pandey$^1$\thanks{biswap@visva-bharati.ac.in} and Rishi
        Khatri$^2$\thanks{khatri@theory.tifr.res.in} \\$^1$
        Department of Physics, Visva-Bharati University, Santiniketan,
        Birbhum, 731235, India \\ $^2$ Department of Theoretical Physics, Tata Institute of Fundamental
        Research, Mumbai, 400005, India}

 \date{\today}

 \pubyear{2018}
  
\begin{document}
\label{firstpage}
\pagerange{\pageref{firstpage}--\pageref{lastpage}}      
\maketitle
       
\begin{abstract}
We analyze two volume limited galaxy samples from the SDSS photometric
and spectroscopic data to test the isotropy in the local Universe. We
use information entropy to quantify the global anisotropy in the
galaxy distribution at different length scales and find that the
galaxy distribution is highly anisotropic on small scales. The
observed anisotropy diminishes with increasing length scales and
nearly plateaus out beyond a length scale of $200 \hmpc$ in both the
datasets. We compare these anisotropies with those predicted by the
mock catalogues from the N-body simulations of the $\Lambda$CDM model
and find a fairly good agreement with the observations. We find a
small residual anisotropy on large scales which decays in a way that
is consistent with the linear perturbation theory. The slopes of the
observed anisotropy converge to the slopes predicted by the linear
theory beyond a length scale of $\sim 200 \hmpc$ indicating a
transition to isotropy. We separately compare the anisotropies
observed across the different parts of the sky and find no evidence
for a preferred direction in the galaxy distribution.
  
\end{abstract}

       
       \section{Introduction}

The statistical homogeneity and isotropy of the Universe is a
fundamental premise of modern cosmology. The assumption of homogeneity
and isotropy of the Universe is known as the Cosmological principle.
The isotropy of the Universe demands that at any given instant of
time, there are no special directions in the Universe. This
assumption, considered to be a cornerstone of the standard
cosmological model, can be tested with the plethora of data from the
modern cosmological observations. The discovery of the Cosmic
Microwave Background radiation (CMB) in 1964 \citep{penzias} proved to
be a milestone in the establishment of the standard model of
cosmology. The Cosmic Background Explorer (COBE) mission launched in
1990 revealed the uniformity of the CMB temperature across the entire
sky \citep{smoot,fixsen} and provided one of the most powerful
evidences in favour of isotropy. COBE and two subsequent space
missions, Wilkinson Microwave Anisotropy Probe (WMAP) and Planck,
along with numerous ground and balloon based experiments
\citep{wmap,adeplanck3,boomerang,act,spt} measured the small
anisotropies in the CMB sky revealing a wealth of cosmological
information and ushering in the era of precision cosmology. The CMB
anisotropies are small fluctuations in the CMB temperature or
intensity as a function of position on the sky.

We should clarify here that even though there are small anisotropies
in the CMB, we can still have \emph{statistical isotropy}. Statistical
isotropy implies that the small fluctuations in the CMB should look
same in every direction in the sky. In other words, the angular power
spectrum of anisotropies should not depend on which part of the sky we
choose to study. Even on the largest scales observable today and super
horizon scales, we should expect small anisotropies and
inhomogeneities of 1 part in $10^4-10^5$ seeded during inflation. We
are however interested in \emph{anomalous anisotropy}, defined as the
anisotropy that is much larger than $10^{-4}-10^{-5}$ expected from
the simplest models of inflation. In this paper we will use word
isotropy to mean isotropy in this statistical sense on small scales
and absence of anomalously large anisotropy on large scales and not
absolute isotropy. The assumption of isotropy and homogeneity in this
sense in the cosmological model allows us to treat the small
anisotropies and inhomogeneities perturbatively on an absolute
isotropic and homogeneous FRW background Universe.

A large number of studies from WMAP and PLANCK \citep{eriksen,
  hoftuft,
  akrami,adeplanck1,adeplanck2,schwarz1,land,hanlewis,moss,grupp,dai}
reveal subtle anomalies in the CMB anisotropies that might challenge
the assumption of \emph{statistical isotropy}.  Several unexpected
features at large angular scales such as a hemispherical power
asymmetry and an abnormally large cold spot, although modestly
significant, suggest a critical examination of the assumption of
isotropy. The assumption of isotropy needs to be tested with
independent datasets and with diverse statistical measures.  This has
been done with various other observations such as the X-ray background
\citep{wu,scharf}, radio sources \citep{wilson,blake}, Gamma-ray
bursts \citep{meegan,briggs}, supernovae \citep{gupta,lin}, galaxies
\citep{marinoni, gibel, yoon, alonso, pandey17}, galaxy clusters
\citep{bengaly17} and neutral hydrogen \citep{hazra}. All these
observations are consistent with the assumption of statistical
isotropy. Contrary to these findings, there are also other studies
with Type-Ia supernovae
\citep{schwarz2,campanelli,kalus,javanmardi,bengaly}, radio sources
\citep{jackson}, galaxies \citep{javanmardi17}, galaxy luminosity
function \citep{appleby} and large scale bulk flows \citep{watkins,
  kashlinsky1, kashlinsky2} which reported statistically significant
deviation from isotropy.  The current observational status does not
provide a clear consensus on the issue of isotropy of the Universe on
large scales and further investigations are necessary to either
establish or refute it.

The large-scale structures in the Universe emerge from the
gravitational collapse of the primordial density fluctuations. The
gravitational collapse is anisotropic in nature. In the Zeldovich
approximation, an overdense region in the dark matter distribution
first collapses along its shortest axis leading to a sheet-like
structure \citep{zeldovich1970}. The subsequent collapse along the medium and the longest
axis results in an elongated filament and a dense compact cluster
respectively \citep{sz1989}. Galaxies are a biased tracer of the underlying mass
distribution and any anisotropy in the distribution of the dark matter
is also expected to be present in the galaxy distribution. The modern
redshift surveys like 2dFGRS \citep{colles} and SDSS \citep{york} have
now mapped a large number galaxies in the local Universe providing an
unprecedented view of the galaxy distribution in our
neighbourhood. The galaxies are found to be distributed in an
interconnected network of filaments, sheets and clusters which are
surrounded by large empty regions. The filaments, which acts as
interconnecting bridges between the clusters are known to be
statistically significant up to $\sim 70-80 \hmpc$
\citep{bharad04,pandey05}. The Sloan Great Wall discovered in the SDSS
\citep{gott05} is one of the richest galaxy system in the nearby
Universe and appears to be contiguous over length scales of more
than 400 Mpc. Observations suggests that there are voids of enormous
sizes such as the Bootes void with radius of 62 Mpc \citep{kirshner}
and the Eridanus supervoid extending up to 300 Mpc
\citep{szapudi}. The Eridanus void is also known to be aligned with
the CMB cold spot and believed to be associated with it. Observational
detection of all these gigantic cosmic structures re-emphasize the
necessity of testing the isotropy in the galaxy distribution.

The SDSS is a multiband photometric and spectroscopic redshift survey
which covers one quarter of the celestial sphere in the Northern
Galactic Cap. The photometric and spectroscopic catalogues of the SDSS
now provide redshifts of millions of galaxies making them suitable for
testing isotropy in the galaxy distribution. The spectroscopic
redshifts are estimated from the spectra of galaxies and hence they
are more reliable but difficult and costly to measure for a very large
number of galaxies. On the other hand, there is a larger uncertainty
in the estimate of the photometric redshifts but photometric data is
easier to obtain for a large number of galaxies. Keeping this in mind,
we consider both the spectroscopic and photometric catalogues from the
SDSS for the present analysis. 

Information entropy can be used to test
the homogeneity and isotropy of the Universe \citep{pandey13,
  pandey16a}. In this work, we use an information theory based method
\citep{pandey16a} to test the isotropy of the local Universe using the
spectroscopic and photometric redshift catalogues of the Sloan Digital
Sky Survey (SDSS).

A brief outline of the paper follows. We describe the method of
analysis in Section 2, the data in Section 3 and present the
results and conclusions in Section 4 and Section 5 respectively.

We have used a $\Lambda$CDM cosmological model with matter density
parameter today, the $\Omega_{m0}=0.31$, dark energy density parameter
$\Omega_{\Lambda0}=0.69$ and Hubble parameter $h=1$ for converting
redshifts to distances throughout the analysis.


\section{METHOD OF ANALYSIS}

We use the anisotropy parameter defined in \citet{pandey16a} to
quantify the anisotropy in the galaxy distribution. This anisotropy
parameter uses the information entropy \citep{shannon48} to measure
the non-uniformity in the distribution of galaxies.  In this method,
we first need to divide the entire sky into pixels of equal area and
similar shape. The equal area of the pixels ensures that the solid
angle subtended by each of these pixels on the observer is the same
whereas the similar shapes ensure the same geometry for each of the
angular bins. We use the Hierarchical Equal Area isoLatitude
Pixelization (HEALPix) software \citep{gorski1,gorski2} for this
purpose. We use the HEALPix resolution parameter $N_{\rm side}$ to
pixelate the sky into $N_{\rm pix}=12 \times N_{\rm side}^2$ pixels of
equal area. The angular size of each pixel for a specific choice of
$N_{\rm side}$ is $\sqrt{\frac{41253}{N_{\rm pix}}}$ degree where
41253 square degree is the total area of the sky. The angular bins
subtended by each of these pixels have the same volumes but may
contain different number of galaxies within them. We choose an upper
limit $r_{\rm max}$ for the radial distance as the galaxies are mapped
only within a finite region. Furthermore, the galaxy surveys very
often do not provide a full sky coverage. The fact that only a part of
the sky is mapped by the survey needs to be taken into account through
a mask specific to the survey. The effective number of pixels $\neff$
available for the analysis is smaller than $N_{\rm pix}$ and depends
on the size and geometry of the mask or the sky coverage of the
survey.

We vary the radial distance within the limit $r_{\rm max}$ with
uniform steps and cumulatively count the number of galaxies within
each angular bin. We use an uniform step size of $10 \hmpc$ throughout
this analysis. We consider a randomly selected galaxy lying within a
radial distance $r$ from the observer. This galaxy resides in any one
of the $\neff$ angular bins and the probability of finding the galaxy
in any particular bin is proportional to the total number of galaxies
residing in that bin. If $n_{i}$ is the number of galaxies located in
the $i^{th}$ angular bin then the probability of finding the galaxy in
the $i^{th}$ bin is given by, $f_{i}=\frac{n_{i}}{\sum^{\neff}_{i=1}
  \, n_{i}}$ and $\sum^{\neff}_{i=1} \, f_{i}=1$ by definition.  So
the event of randomly selecting a galaxy has $\neff$ outcomes each
with a different probability $f_{i}$.  The information entropy
associated with this event for a specific radial distance $r$ can be
written as,
\begin{eqnarray}
H_{lb}(r)& = &- \sum^{\neff}_{i=1} \, f_{i}\, \log\, f_{i} \nonumber\\ &=& 
\log N - \frac {\sum^{\neff}_{i=1} \, n_i \, \log n_i}{N},
\label{eq:shannon2}
\end{eqnarray}
where $N$ is the total number of galaxies located within a radial
distance $r$ from the observer. The subscript $lb$ in $H_{lb}$ implies
that we consider the pixels in the longitude-latitude $(l,b)$ space
and study it as a function of the radial distance $r$ up to which the
number counts are integrated. The base of the logarithm is arbitrary
and only decides the unit of information. We use the base of $10$ for
the present work. If the probabilities $f_{i}$ are identical for each
of the angular bins then the information entropy will be maximum,  $(H_{lb})_{max}=\log \, \neff$. This
would only occur if each of the angular bins hosts exactly the same
number of galaxies within a distance $r$. This corresponds to the
situation when there is maximum uncertainty about the location of the
randomly selected galaxy. We define the anisotropy parameter,
\begin{eqnarray}
a_{lb}(r)=1-\frac{H_{lb}(r)}{\left(H_{lb}\right)_{max}},
\end{eqnarray}
 to measure the degree of radial anisotropy present in any
 distribution. It may be noted that for a completely isotropic
 distribution, the probability distribution is uniform leading to
 $H_{lb}(r)=\left(H_{lb}\right)_{max}$ and $a_{lb}(r)=0$. On the other
 hand, if all the galaxies are located only in one specific bin out of
 the $\neff$ angular bins then the galaxy distribution is maximally
 anisotropic. In this case, there is no uncertainty about the location
 of the randomly selected galaxy and consequently we have
 $H_{lb}(r)=0$ and $a_{lb}(r)=1$.

The galaxies are not distributed randomly but
in a web-like network. The presence of the coherent patterns in the
galaxy distribution like filaments, sheets, clusters and voids cause
the distribution to be highly anisotropic on small scales. So the
probabilities of locating randomly selected galaxies in different
angular bins are not the same. The anisotropy parameter $a_{lb}(r)$
will measure non-uniformity in the distribution of galaxies as a
function of the length scale $r$. We will measure the anisotropy
parameter $a_{lb}(r)$ in the SDSS data as a function of $r$ up to the
maximum radial distance $r_{\rm max}$. If the assumption of isotropy
on large scales holds in the real Universe then we expect the
anisotropy parameter to decrease with the increasing length scale $r$
and should become negligibly small on the scales where the Universe
becomes isotropic.  It should be noted that the value of $a_{lb}(r)$
is also expected to be sensitive to the choice of $N_{\rm side}$ as it
decides the total number of pixels $N_{\rm pix}$. The pixel sizes will
be larger for a smaller $N_{\rm side}$. As a result the volume covered
by each angular bin will be also larger. This would increase the
galaxy counts and reduce the Poisson noise leading to a decrease in
the anisotropy. Even an isotropic distribution of finite size will
exhibit some anisotropy due to the Poisson noise on small scales. To
asses this, we compare our results to that obtained from a homogeneous
and isotropic Poisson distribution which has the same geometry and
sampling density as the actual data.

The anisotropy parameter measured at each length scale for the entire
survey region provides the degree of global anisotropy present in the
galaxy distribution. Besides the global isotropy, it is also important
to compare the degree of anisotropy observed along the different
directions in the sky. For this, we will be dividing the SDSS survey
area into a small number of regions and separately measure the entropy
and anisotropy in each of these regions. This would allow us to test
the statistical isotropy of the galaxy distribution and identify the
existence of any preferred directions.

The information entropy is related to the higher order moments of a
distribution \citep{pandey16b}, in addition to the second order moment
or the power spectrum. The higher order moments of the galaxy density
field are expected to be non zero as the present day galaxy
distribution is known to be highly non-Gaussian. This provides the
argument for using the information entropy as an effective measure of
the anisotropy present in the galaxy distribution, since it captures
information beyond the 2-point correlation function or the power
spectrum.

\begin{figure*}
\resizebox{18 cm}{!}{\rotatebox{0}{\includegraphics{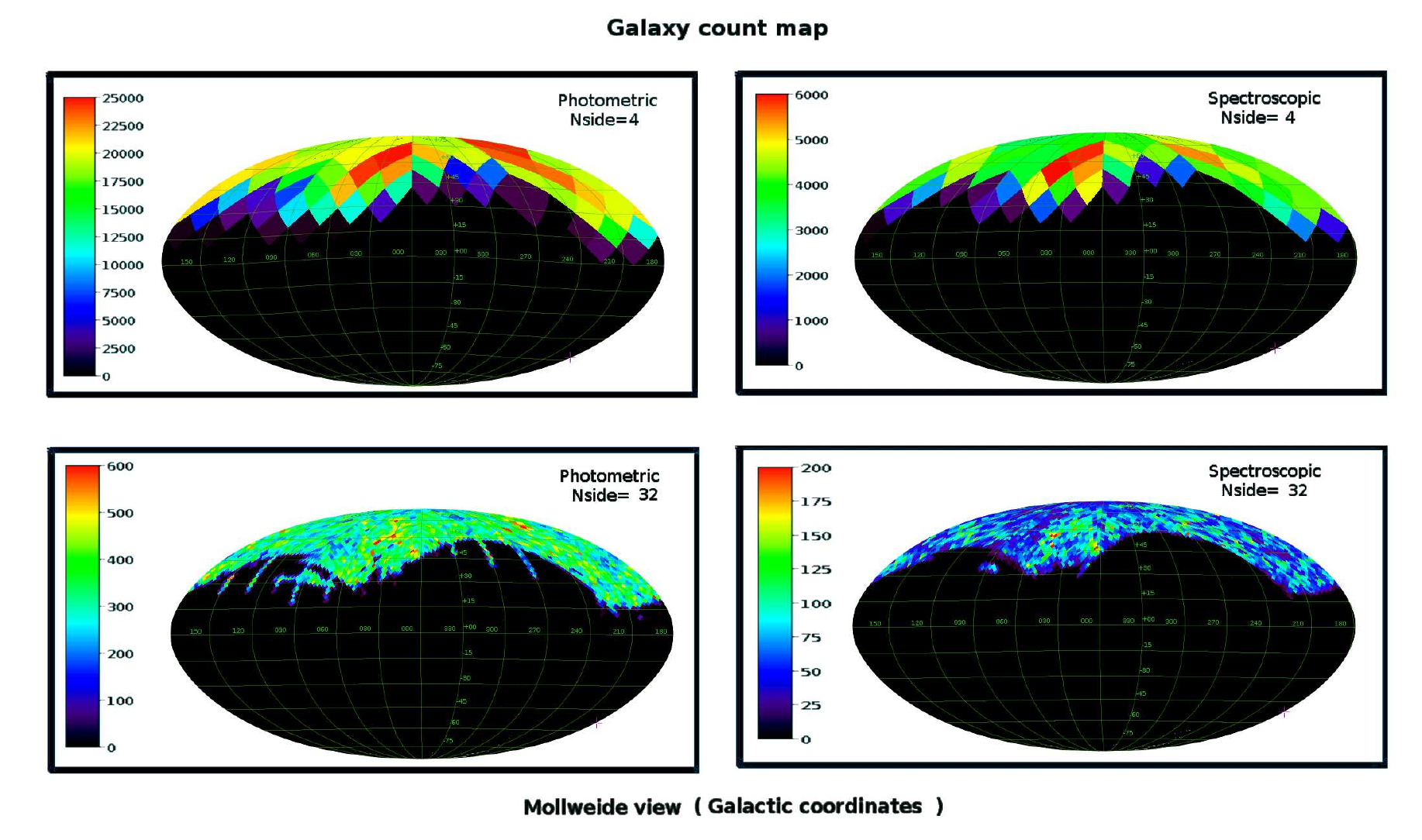}}}\\
\caption{The galaxy number counts for the SDSS photometric and
  spectroscopic samples for two different resolutions, HEALPix $N_{\rm
    side}=4$ and $32$. }
  \label{fig:count_gal}
\end{figure*}

\section{DATA}
We use the data from the twelfth data release of the Sloan Digital Sky
Survey (SDSS, DR12) \citep{alam} which is the final data release of
the SDSS-III.  A description of the telescope used in the SDSS is
provided in \citet{gunn2} and the SDSS camera and filters are
discussed in \citet{gunn1} and \citet{fukugita} respectively. The
target selection algorithm of the Main Galaxy Sample is described in
\citet{strauss}.

\subsection{THE SDSS PHOTOMETRIC DATA}
We use the photometric redshift catalogue for the SDSS data prepared
by \citet{beck}. We select only the galaxies with the photometric
error classes $1$ and $-1$ which have accurate redshift error
estimates. For other error classes, the redshift estimation errors are
dependent on the position in color and magnitude space and hence
requires additional statistical errors to be taken into account. We
apply a cut in the redshift error estimate $\delta_{z}^{\rm
  photo}<0.03$ and consider all the galaxies in the redshift range
$0<z<0.3$ with r-band extinction $A_r<0.18$ which provides us with a
sample containing $2827248$ galaxies. Further cuts in the r-band
apparent magnitude $13<m_r<19$ and r-band absolute magnitude
$M_{r}<-20.4$ are applied to construct a volume limited sample of
galaxies. The resulting volume limited sample of galaxies extends up
to redshift $z<0.2143$ or comoving distance $<609.197 \hmpc$ and
consists of $1022630$ galaxies. The volume limited sample prepared
from the SDSS photometric data does not start from $z=0$ but from
$z=0.017$ which corresponds to a distance of $51.4 \hmpc$ due the
apparent magnitude cut adopted here. We consider only the galaxies in
the Northern Galactic Cap to get a contiguous sky coverage which
reduces the available galaxies to $784329$ in our volume limited
sample.

\subsection{THE SDSS SPECTROSCOPIC DATA}
We use publicly available spectroscopic data from the SDSS
CasJobs\footnote{http://skyserver.sdss.org/casjobs/}. We retrieve the
spectroscopic information of all the galaxies in the redshift range
$0<z<0.3$ yielding $916633$ galaxies. We then construct a volume
limited sample of galaxies by restricting the r-band Petrosian
magnitude in the range $13<m_{r}<17.77$ and r-band absolute magnitude
to $M_{r}<-20.4$.  The values of k-corrections are also obtained from
the SDSS CasJobs. These cuts yield a volume limited sample of $180181$
galaxies distributed within redshift $z<0.1341$ or comoving distance
$<391.8 \hmpc$. The resulting volume limited sample does not start from
$z=0$ but from $z=0.0167$ due the apparent magnitude range chosen. So
the volume limited sample prepared from the SDSS spectroscopic data
starts at $50 \hmpc$ and extends upto $391.8 \hmpc$. We require a
contiguous region of the sky for our analysis. So we only consider the
galaxies in the Northern Galactic Cap which finally leaves us with a
volume limited galaxy sample comprised of $152860$ galaxies.

\subsection{PREPARING THE MASKS}
We use HEALPix to divide the entire sky into pixels of equal sizes and
count the number of galaxies inside each pixel. The galaxy number
counts in different pixels for the volume limited samples of galaxies
constructed from the spectroscopic and photometric redshift catalogues
are shown in \autoref{fig:count_gal}. The top left and bottom left
panels of \autoref{fig:count_gal} show the number counts in the SDSS
photometric data with $N_{\rm side}=4$ and $N_{\rm side}=32$
respectively. The top right and bottom right panels show the same but
for the SDSS spectroscopic data. The galaxy distributions are
3-dimensional and these counts are the integrated galaxy counts inside
each pixel up to the maximum radial extent of the corresponding galaxy
samples. $N_{\rm side}$ sets the resolution of the map. As a result we
see relatively smaller number counts inside the pixels for $N_{\rm
  side}=32$ than $N_{\rm side}=4$. It may be also noted that the
pixels near the boundary of the survey regions preferentially show
lower number counts. This is related to the fact that only parts of
the boundary pixels lie inside the survey region. These pixels need to
be discarded from any analysis of isotropy and we take this into
account by preparing a mask for each dataset and for each resolution.

To prepare the mask we sub-pixelate each pixel into smaller sub-pixels
using a new variable $N_{\rm side}^{\rm sub}$ which we have taken to
be $N_{\rm side}^{\rm sub}=64$ for the present work. We count galaxies
inside each of the sub-pixels and create a mask map by assigning a
value 1 to the non-empty sub-pixels and 0 to the empty sub-pixels. We
then degrade the mask map to a new mask map for $N_{\rm side}$. The
pixels for which all the sub-pixels are empty have value $0$ and are
masked. The values for pixels at the boundary indicate the fraction
of non-empty sub-pixels in that pixel.  We identify and mask the
sparsely populated pixels near the boundary by using a threshold value
$m_{\rm th}$ for this map. If a pixel in the $N_{\rm side}$ map has a
value $< m_{\rm th} $ then we mask the corresponding pixel. In our
analysis we have used $m_{\rm th} = 0.75$.

\subsection{RANDOM MOCK CATALOGUES}
We generate $10$ Poisson random distributions for both the volume
limited samples constructed from the spectroscopic and photometric
data. The mean density of these samples is chosen to be same as the
respective volume limited samples under consideration. We then apply
the respective masks to generate $10$ random mock catalogues each for
the SDSS spectroscopic and photometric samples.

\subsection{MOCK CATALOGUES FROM N-BODY SIMULATIONS}
We use a Particle-Mesh (PM) N-Body code to simulate the present day
distributions of dark matter in the $\Lambda$CDM model. We use
$\Omega_{m0}=0.31$, $\Omega_{\Lambda0}=0.69$, $h=0.68$,
$\sigma_{8}=0.81$ and $n_{s}=0.96$ \citep{adeplanck3} as the values of
the cosmological parameters. We simulate the distributions using
$256^{3}$ particles on a $512^{3}$ grid in a comoving volume of $(
1433.6 \, h^{-1}\, {\rm Mpc})^3$. We run the simulations for three
different realizations of the initial density fluctuations. We place
an observer inside the centre of each simulation box and map the
distributions to redshift space using the peculiar velocities of
particles. The individual particles are treated as galaxies and we
construct $3$ mock catalogues from each of the three boxes for each of
the two volume limited SDSS samples (photometric and spectroscopic) by
sampling particles with the same mean density as the corresponding
SDSS sample and applying the respective masks. This provides us with
$9$ mock galaxy catalogues for the $\Lambda$CDM model for
each of the volume limited SDSS samples.

\section{RESULTS}

\subsection{THE GLOBAL ANISOTROPY IN THE PHOTOMETRIC AND SPECTROSCOPIC GALAXY SAMPLES}

\begin{figure*}
\resizebox{8
  cm}{!}{\rotatebox{0}{\includegraphics{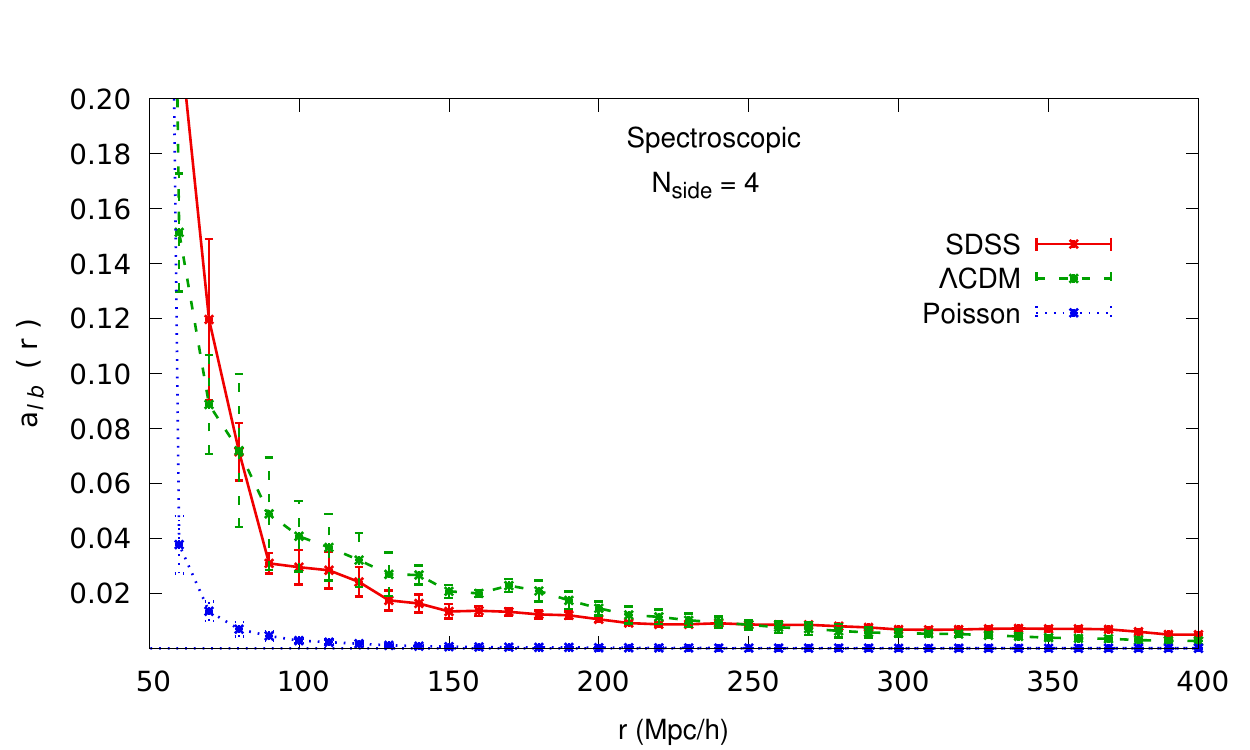}}}
\resizebox{8
  cm}{!}{\rotatebox{0}{\includegraphics{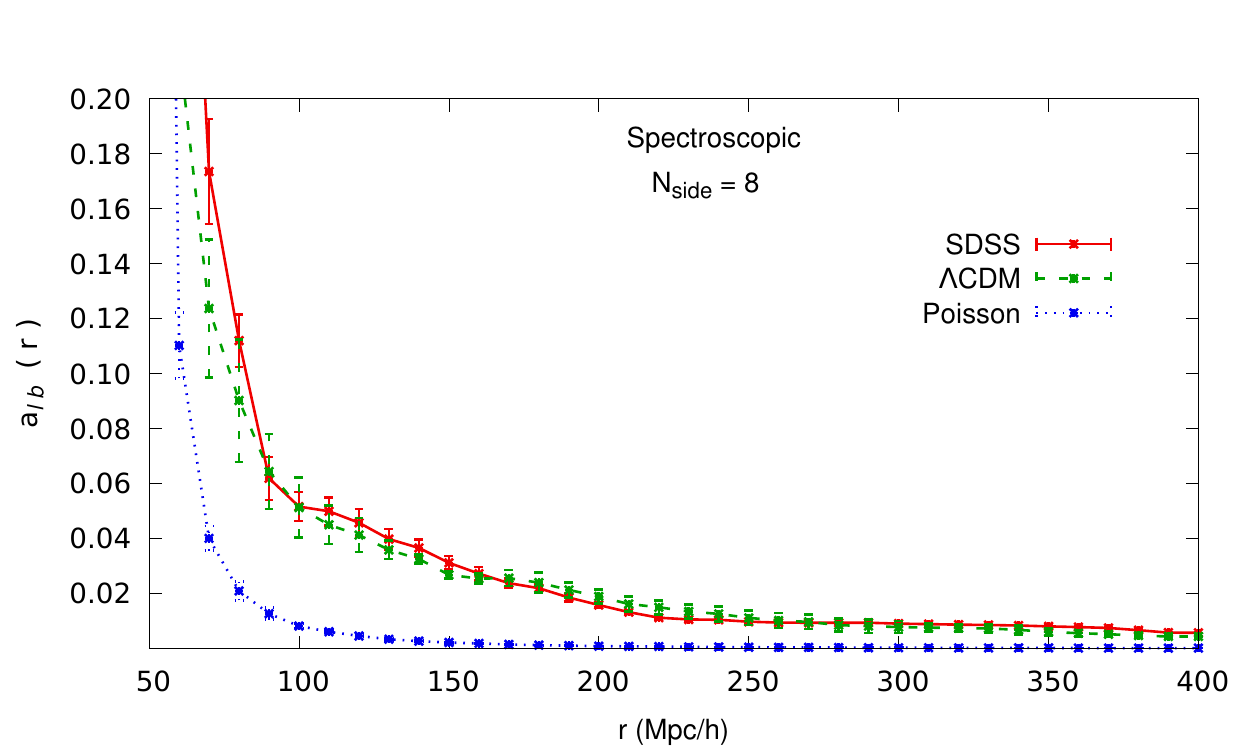}}}\\ \resizebox{8
  cm}{!}{\rotatebox{0}{\includegraphics{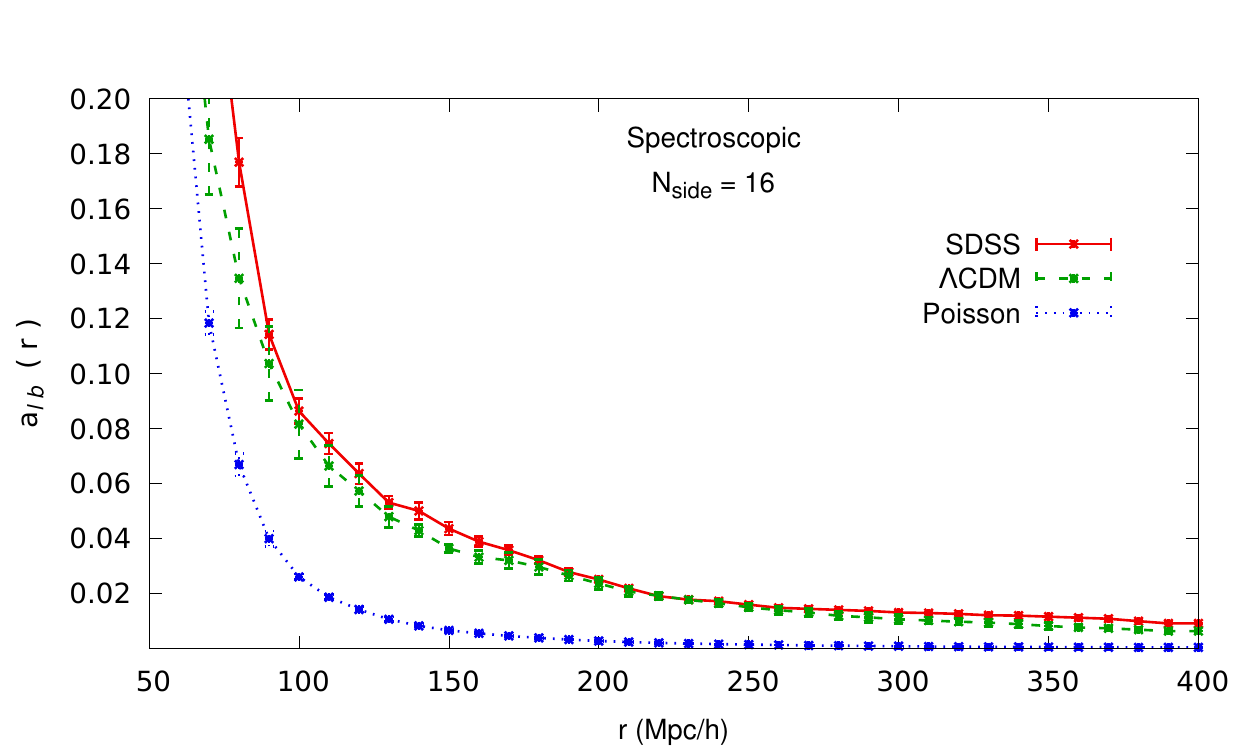}}}
\resizebox{8
  cm}{!}{\rotatebox{0}{\includegraphics{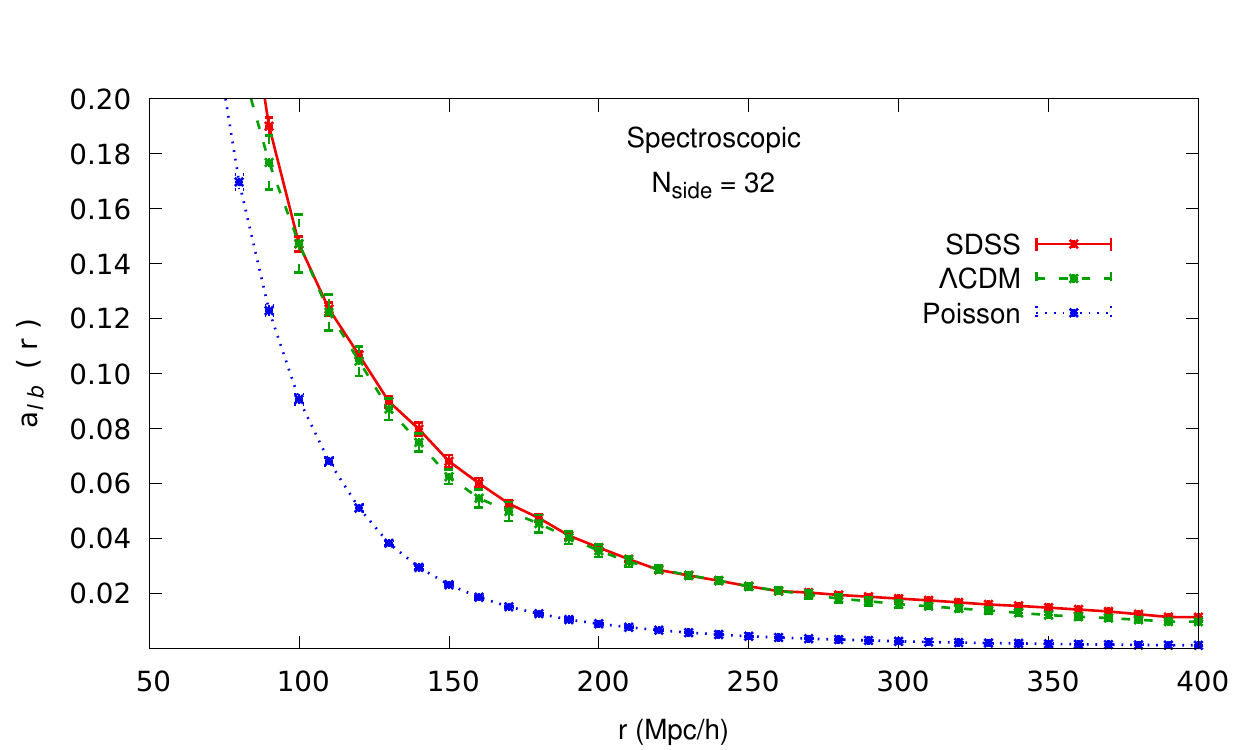}}}\\
\caption{The anisotropy parameter $a_{lb}(r)$ measured as a function
  of radial distance $r$ for SDSS spectroscopic data along with the
  same for the mock galaxy catalogues drawn from the random
  distributions and N-body simulations of the $\Lambda$CDM model. Each
  panel shows the results for a different resolution of the maps
  decided by the value of $N_{\rm side}$ indicated in that
  panel. $1-\sigma$ errorbars shown on each SDSS data points are
  estimated from the $10$ Jackknife samples drawn from the data. The
  $1-\sigma$ errorbars for the $\Lambda$CDM model and Poisson random
  distribution are estimated from $9$ and $10$ independent
  realizations respectively.}
  \label{fig:global_s}
\end{figure*}

\begin{figure*}
\resizebox{8 cm}{!}{\rotatebox{0}{\includegraphics{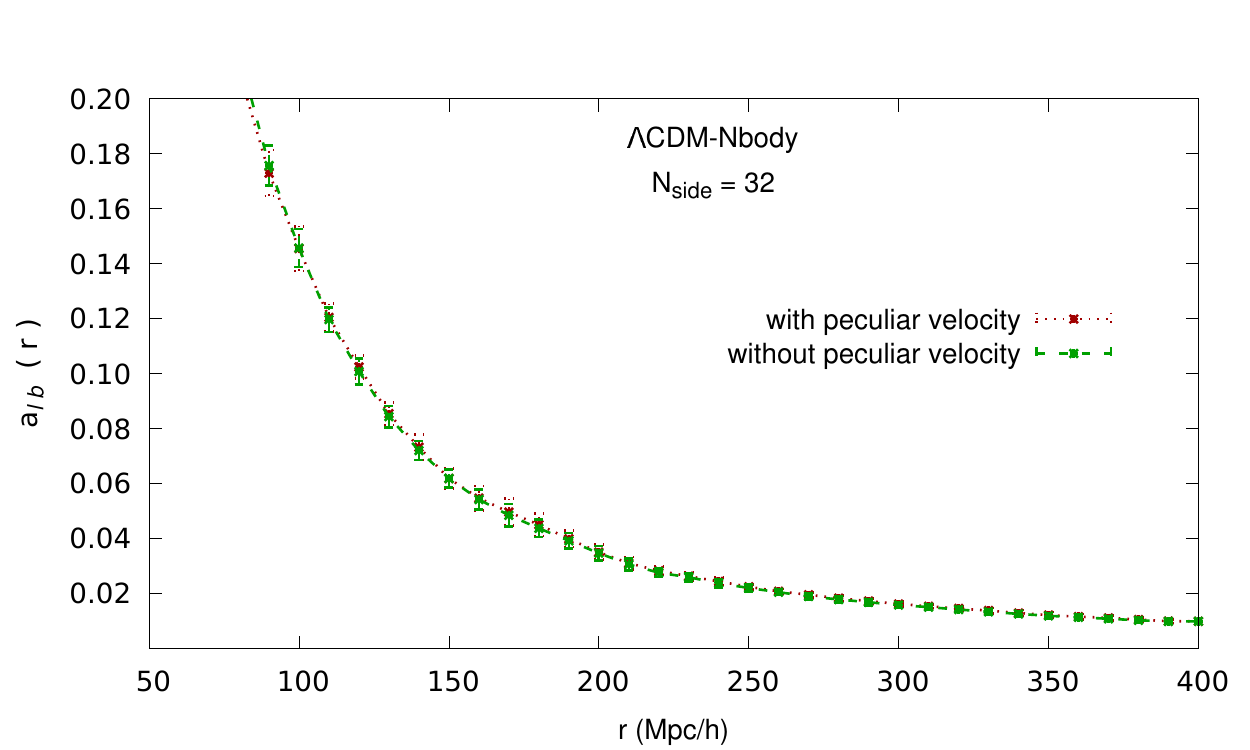}}}\\
\caption{Comparison of the anisotropy parameter $a_{lb}(r)$ of the
  mock galaxy samples when they are prepared with and without taking
  into account the effect of peculiar velocities (redshift space
  distortions). The $1-\sigma$ errorbars shown are estimated using
  $10$ independent realizations.}
  \label{fig:pecvel}
\end{figure*}

\begin{figure*}
\resizebox{8 cm}{!}{\rotatebox{0}{\includegraphics{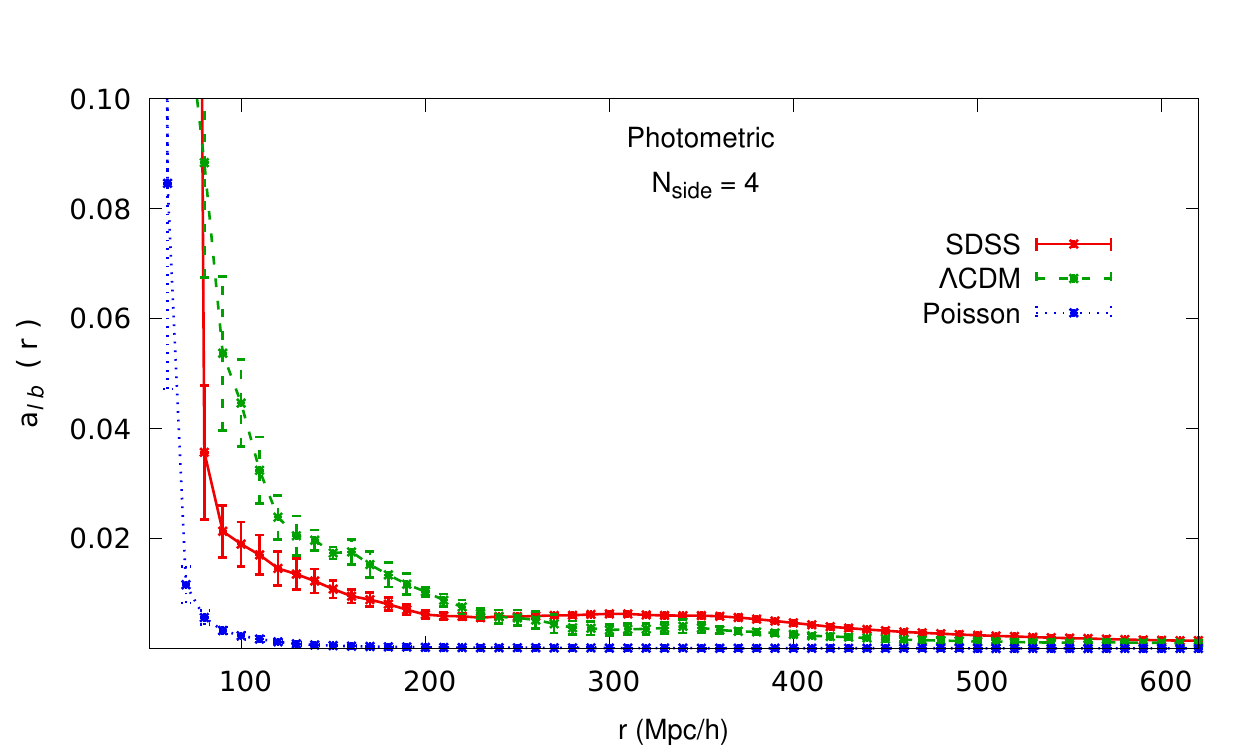}}}
\resizebox{8 cm}{!}{\rotatebox{0}{\includegraphics{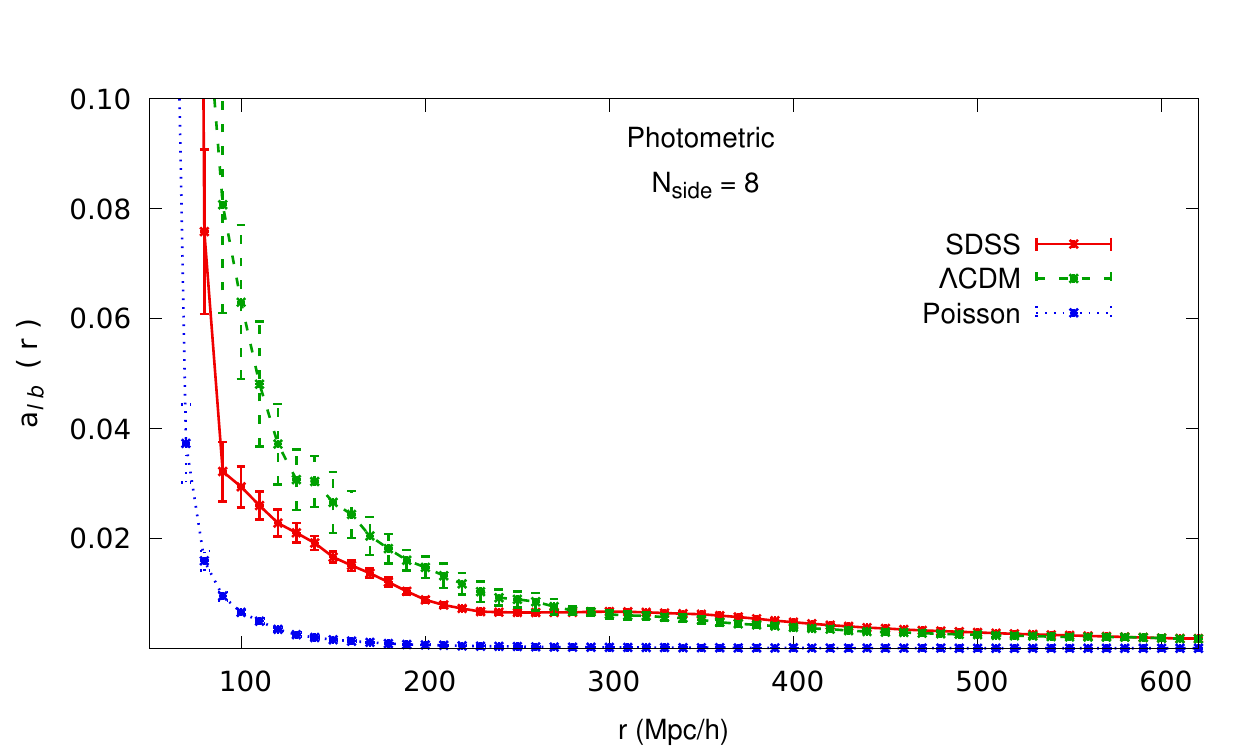}}}\\
\resizebox{8 cm}{!}{\rotatebox{0}{\includegraphics{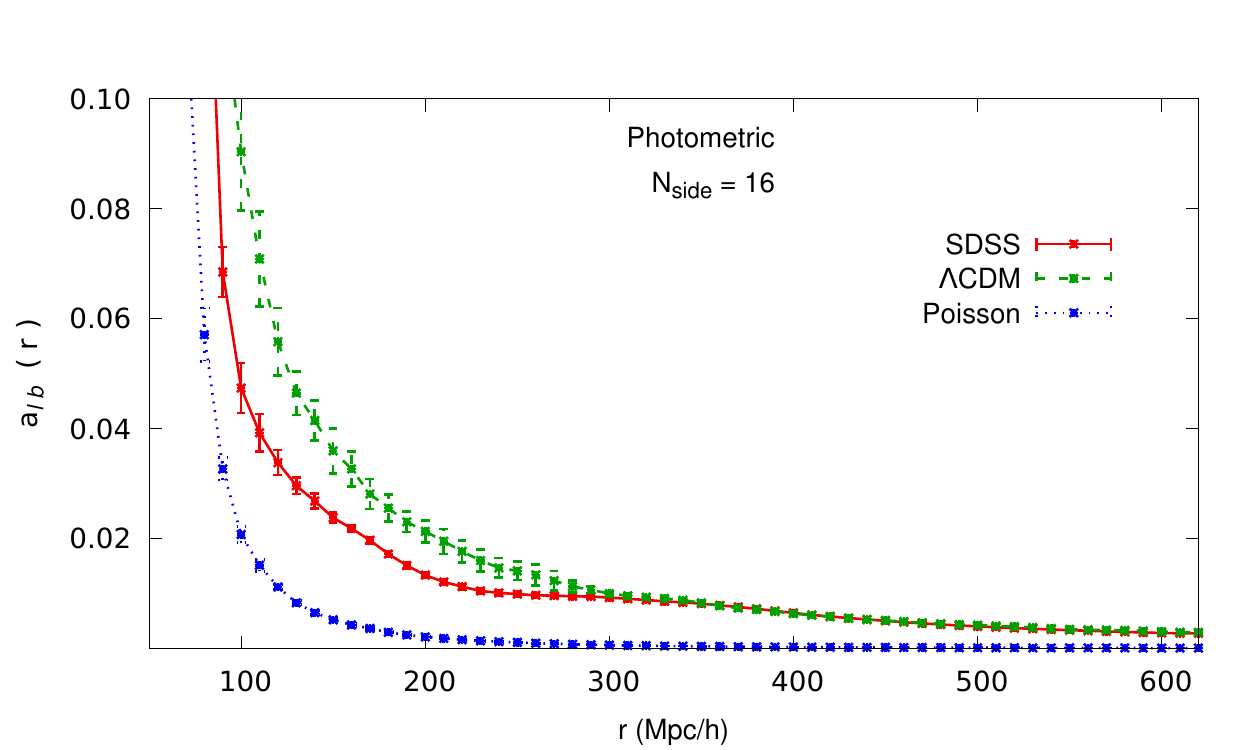}}}
\resizebox{8 cm}{!}{\rotatebox{0}{\includegraphics{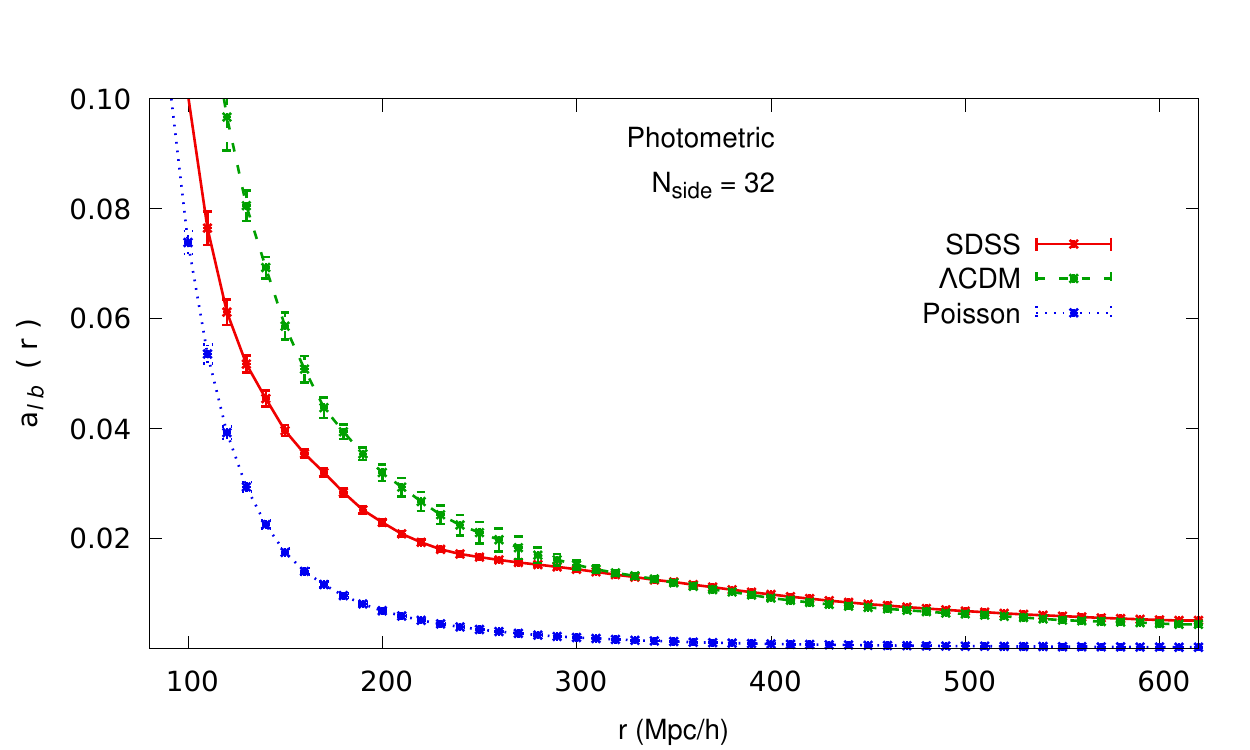}}}\\
\caption{Same as \autoref{fig:global_s} but for the SDSS photometric data.}
  \label{fig:global_p}
\end{figure*}

\begin{figure*}
\resizebox{8 cm}{!}{\rotatebox{0}{\includegraphics{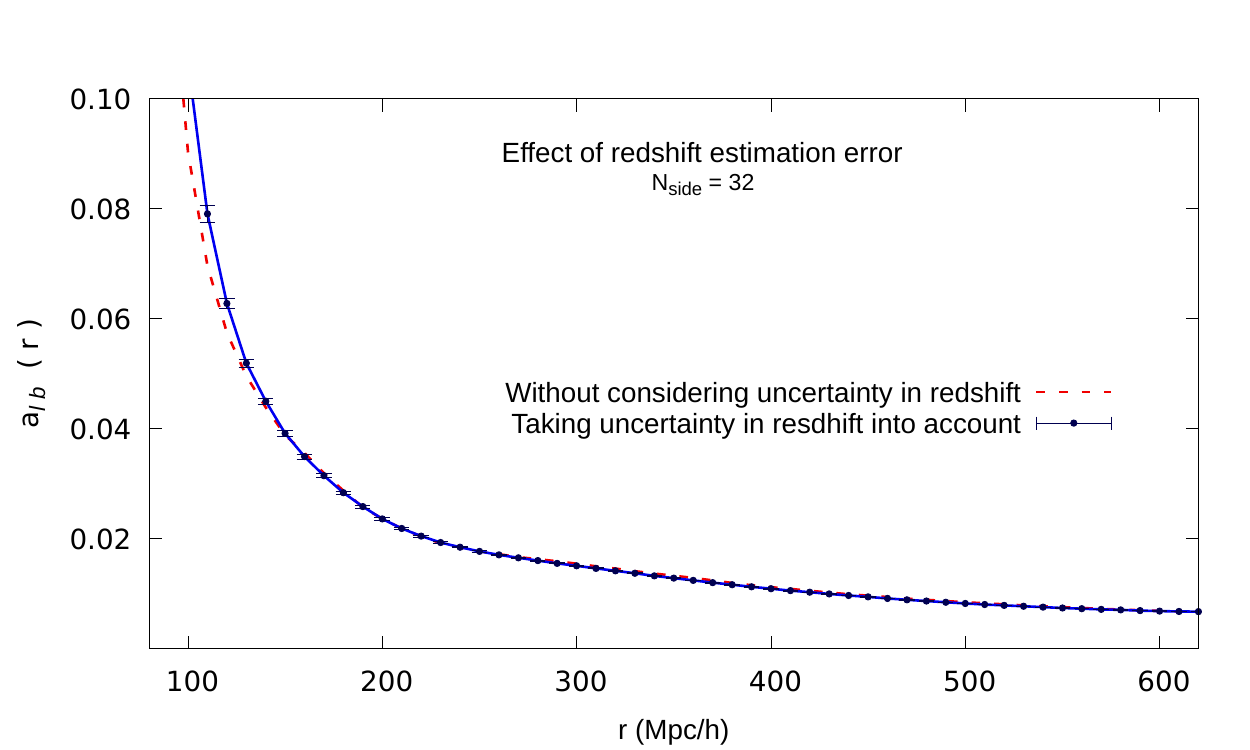}}}\\
\caption{Effect of the uncertainty in the measurements of the
  photometric redshifts \citep{beck} in the SDSS. The $1-\sigma$
  errorbars shown are estimated using $10$ independent
  realizations. The anisotropy parameter $a_{lb}$ is insensitive to
  small errors in the photometric redshift estimates.}
  \label{fig:errz}
\end{figure*}

\begin{figure*}
\resizebox{8 cm}{!}{\rotatebox{0}{\includegraphics{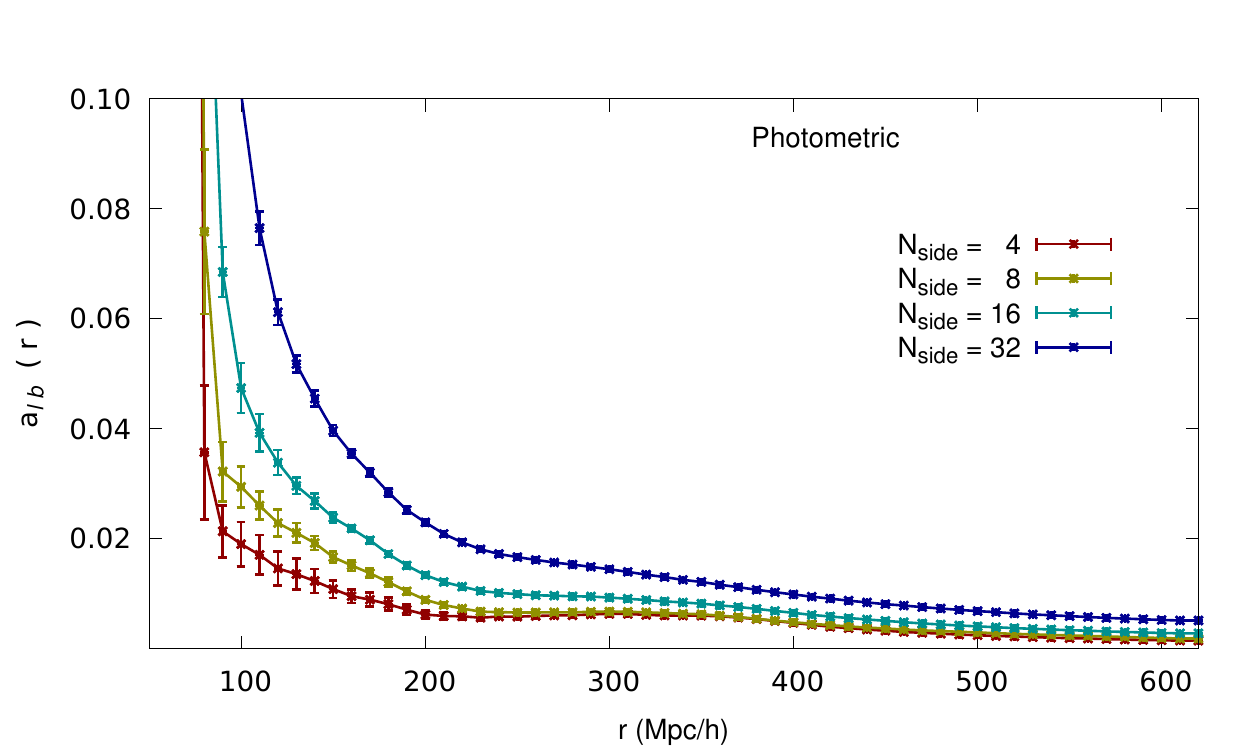}}}
\resizebox{8 cm}{!}{\rotatebox{0}{\includegraphics{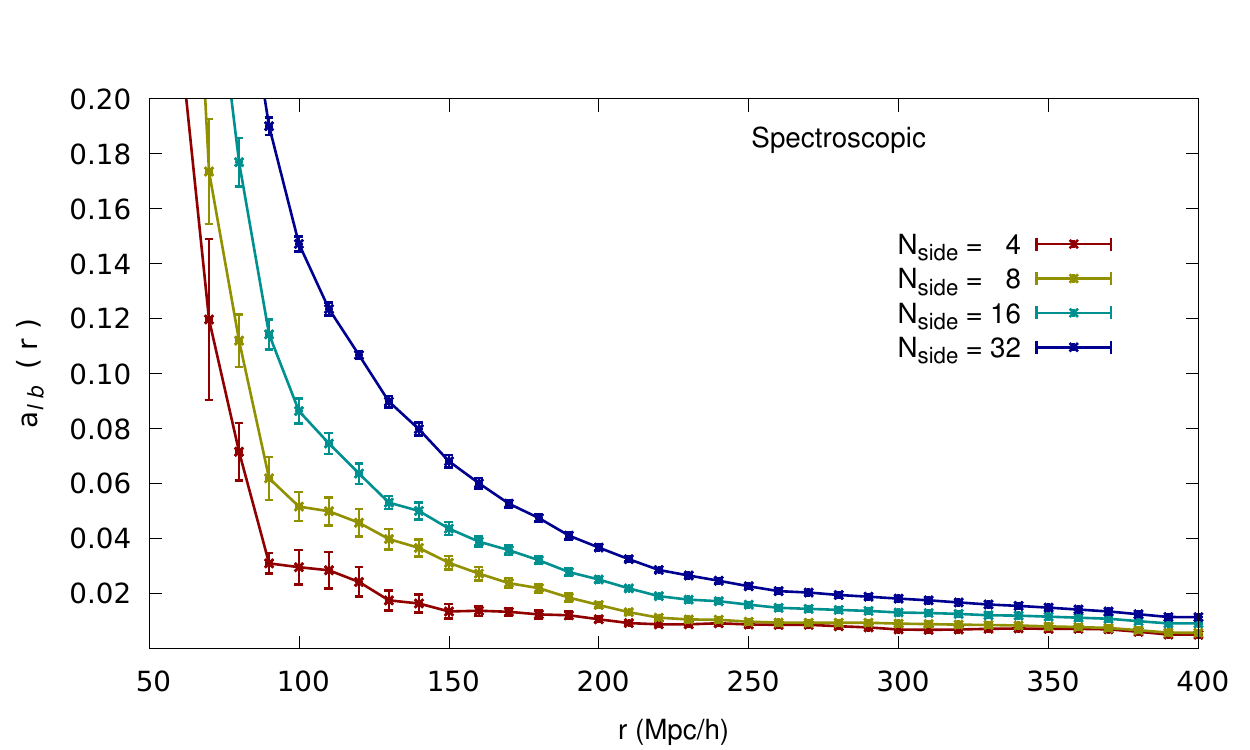}}}\\
\resizebox{8 cm}{!}{\rotatebox{0}{\includegraphics{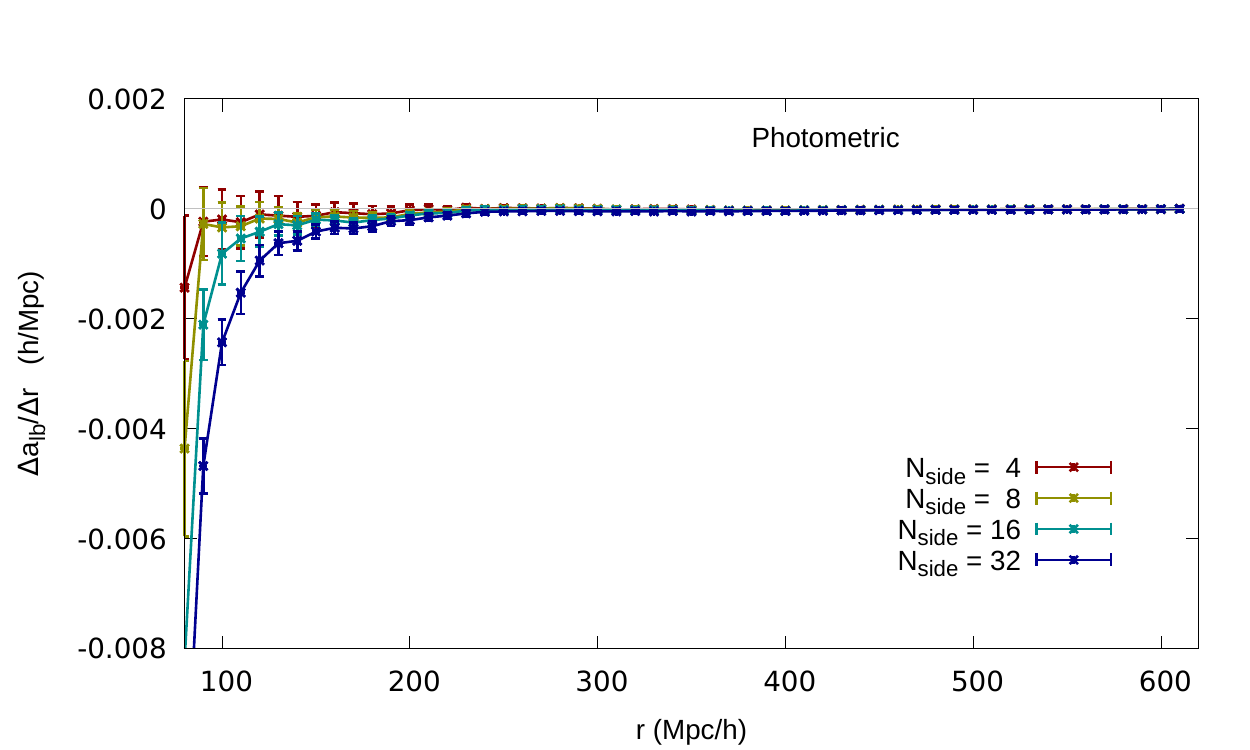}}}
\resizebox{8 cm}{!}{\rotatebox{0}{\includegraphics{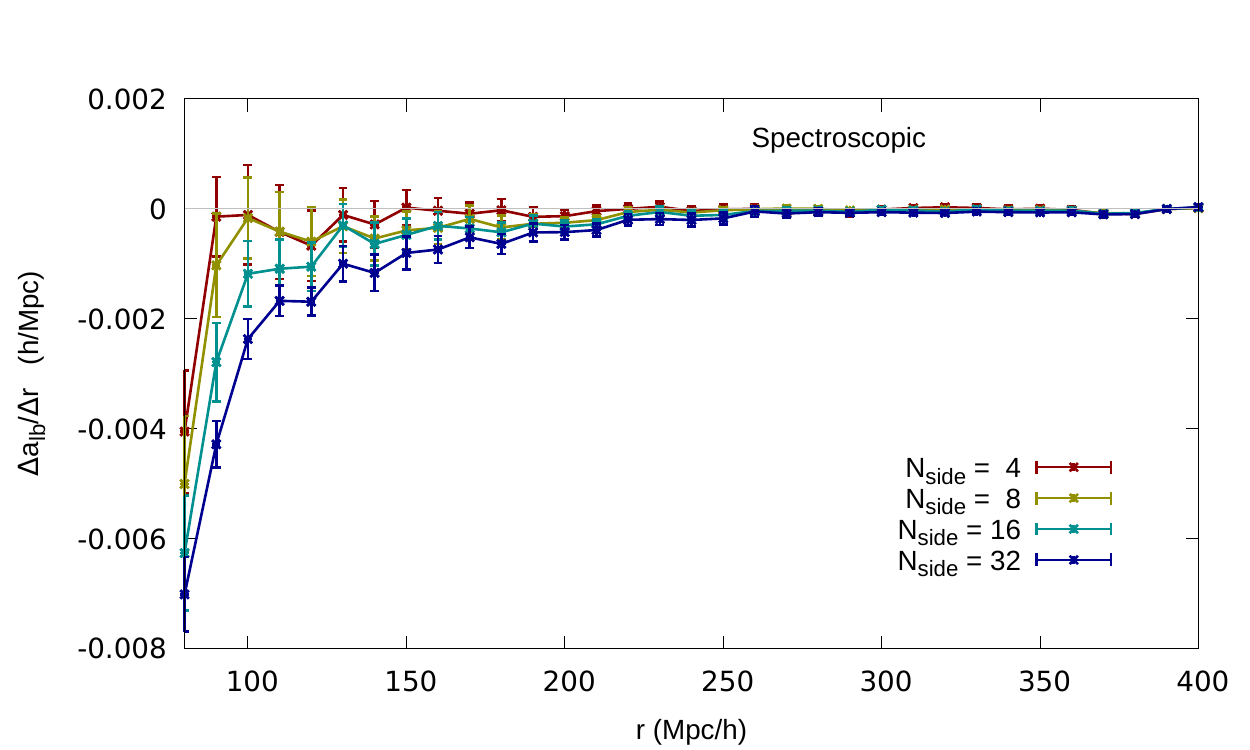}}}\\
\caption{The anisotropy parameter $a_{lb}(r)$ for the SDSS photometric
  and spectroscopic data as a function of radial distance for
  different $N_{\rm side}$ values used for analysis. The bottom panels
  show the rate of change of the anisotropy in each case as measured
  by the slopes of the curves shown in the top panels.}
  \label{fig:global_sp}
\end{figure*}

\begin{figure*}
  \resizebox{8 cm}{!}{\rotatebox{0}{\includegraphics{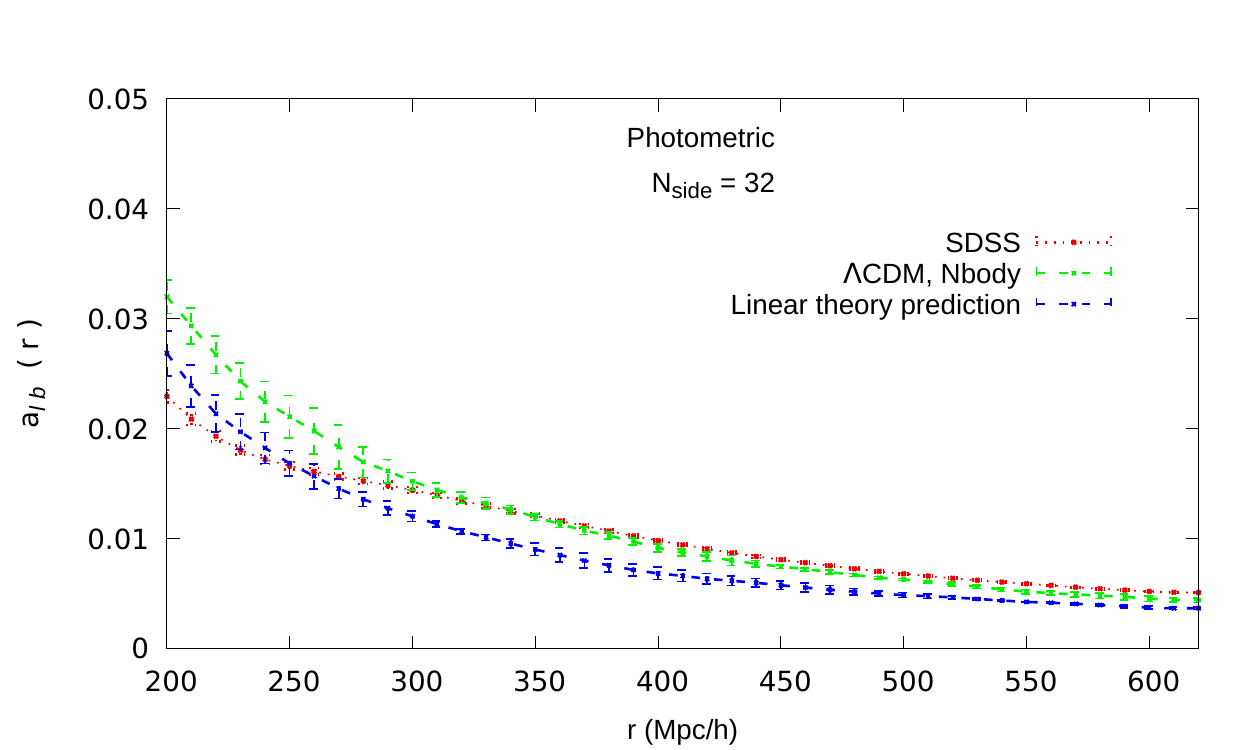}}}
  \resizebox{8 cm}{!}{\rotatebox{0}{\includegraphics{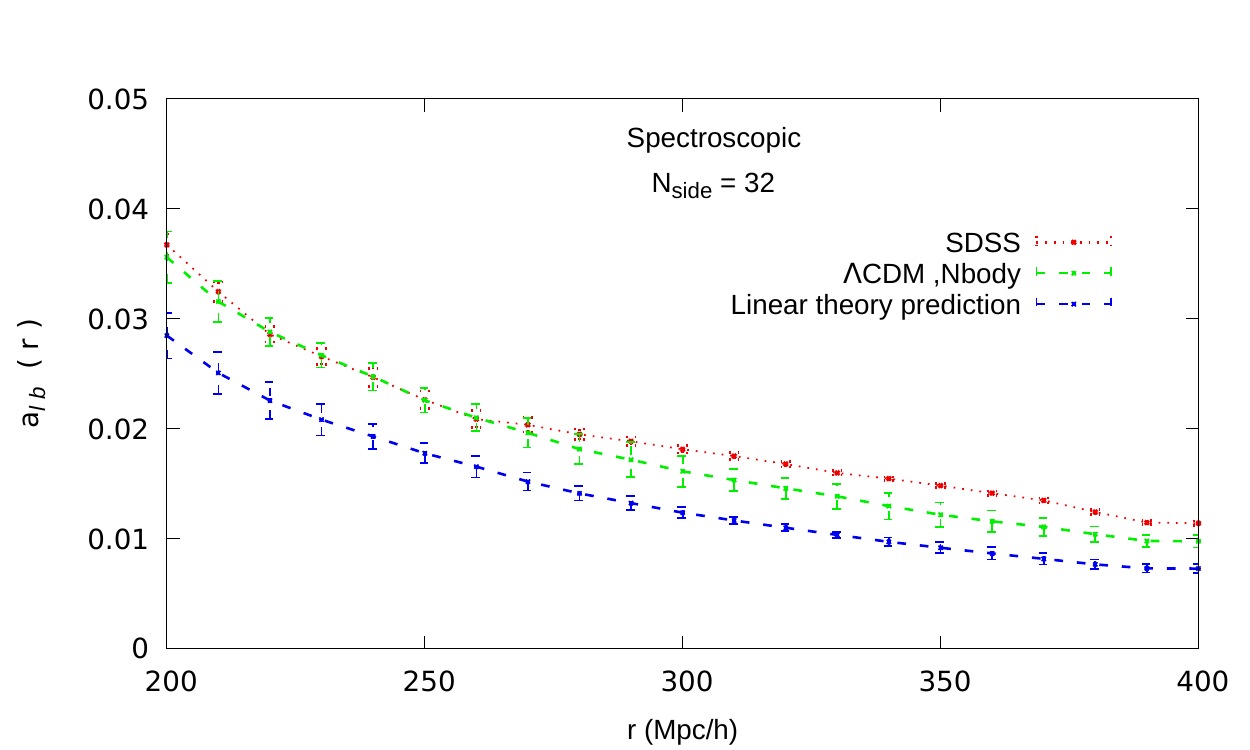}}}\\
  \resizebox{8 cm}{!}{\rotatebox{0}{\includegraphics{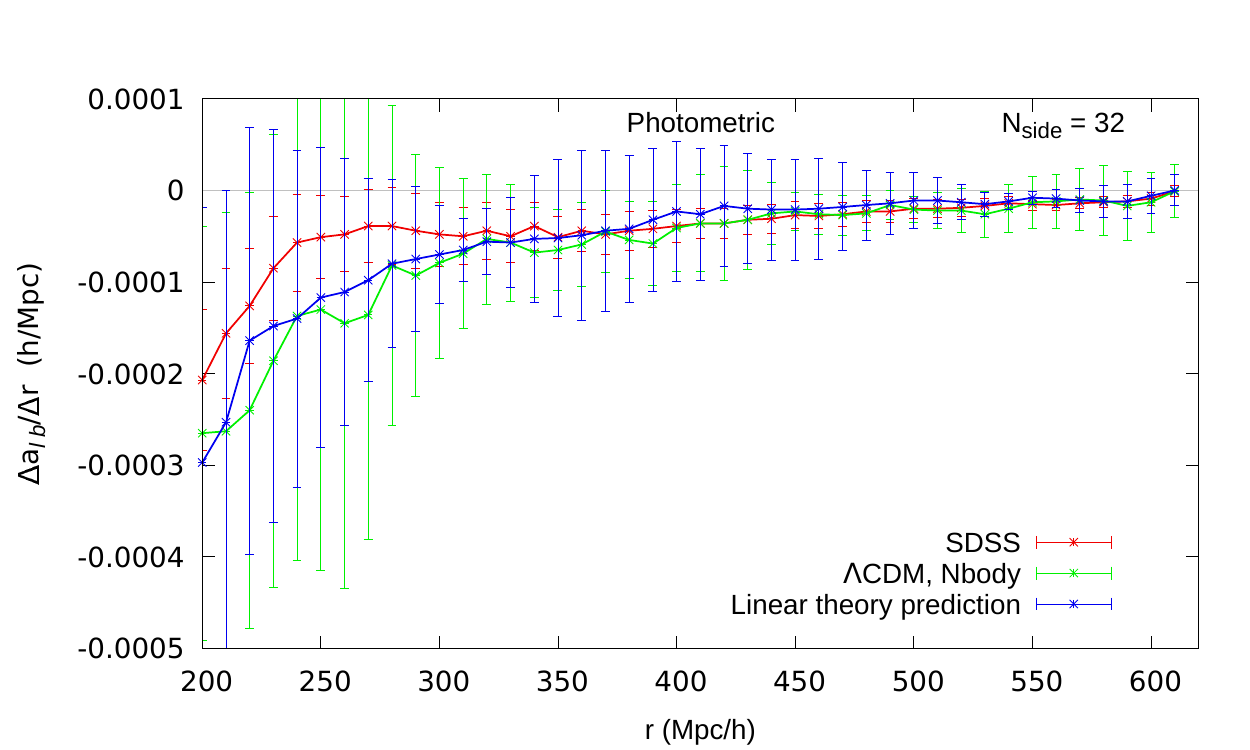}}}
  \resizebox{8 cm}{!}{\rotatebox{0}{\includegraphics{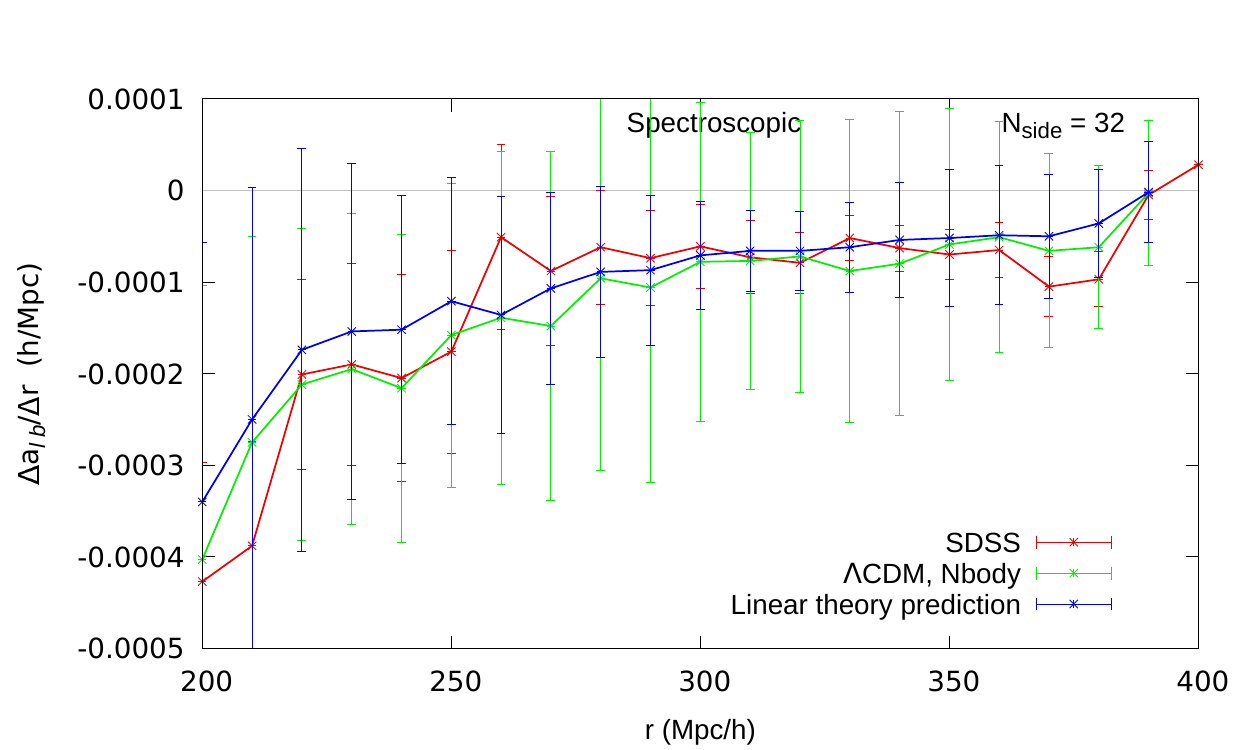}}}\\
  \caption{The top left panel compares the residual anisotropy on
    large scales observed in the SDSS photometric sample, $\Lambda$CDM
    N-body simulations and that expected from the linear theory. The
    top right panel compares the same for the SDSS spectroscopic
    sample. The bottom left panel shows the variation of slopes of the
    anisotropy parameter as a function of length scales for the SDSS
    photometric sample along with the predictions from the $\Lambda$CDM
    N-body simulations and linear theory. The right panel compares the
    same but for the SDSS spectroscopic sample. The $1-\sigma$
    errorbars shown in each case are estimated using $10$ independent
    realizations. }
  \label{fig:slope_lin}
\end{figure*}


\begin{figure*}
\resizebox{18 cm}{!}{\rotatebox{0}{\includegraphics{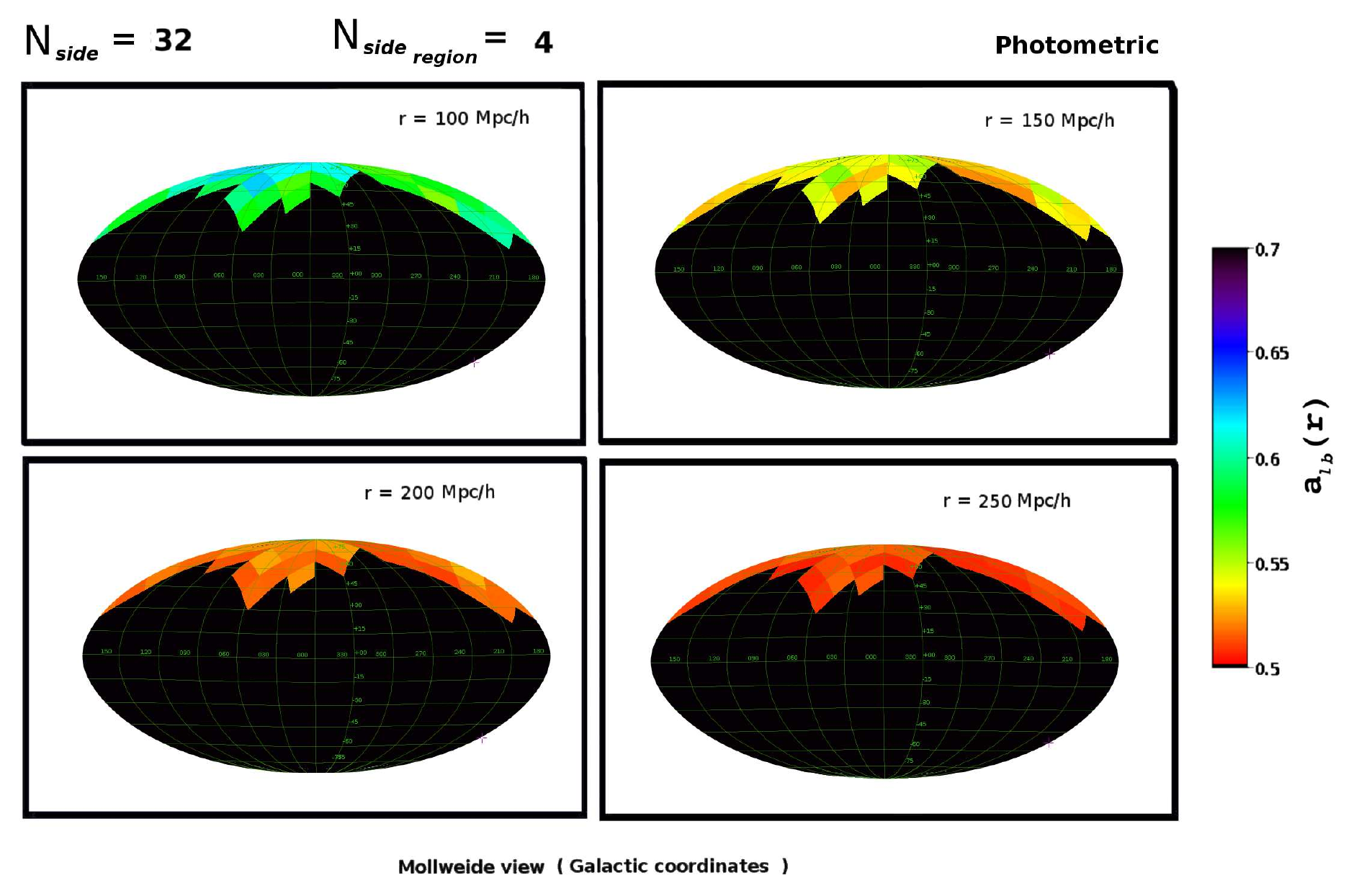}}}\\
\caption{The anisotropy parameter $a_{lb}(r)$ in the $23$ separate
  regions of the SDSS photometric sky at different radii $r$. The
  pixelization is done with $N_{\rm side}=32$ for calculating $a_{lb}$
  in each region, with each region corresponding to a pixel of $N_{\rm
    side}^{\rm region}=4$ HEALPix map.}.
  \label{fig:entropymap_p}
\end{figure*}

\begin{figure*}
\resizebox{18 cm}{!}{\rotatebox{0}{\includegraphics{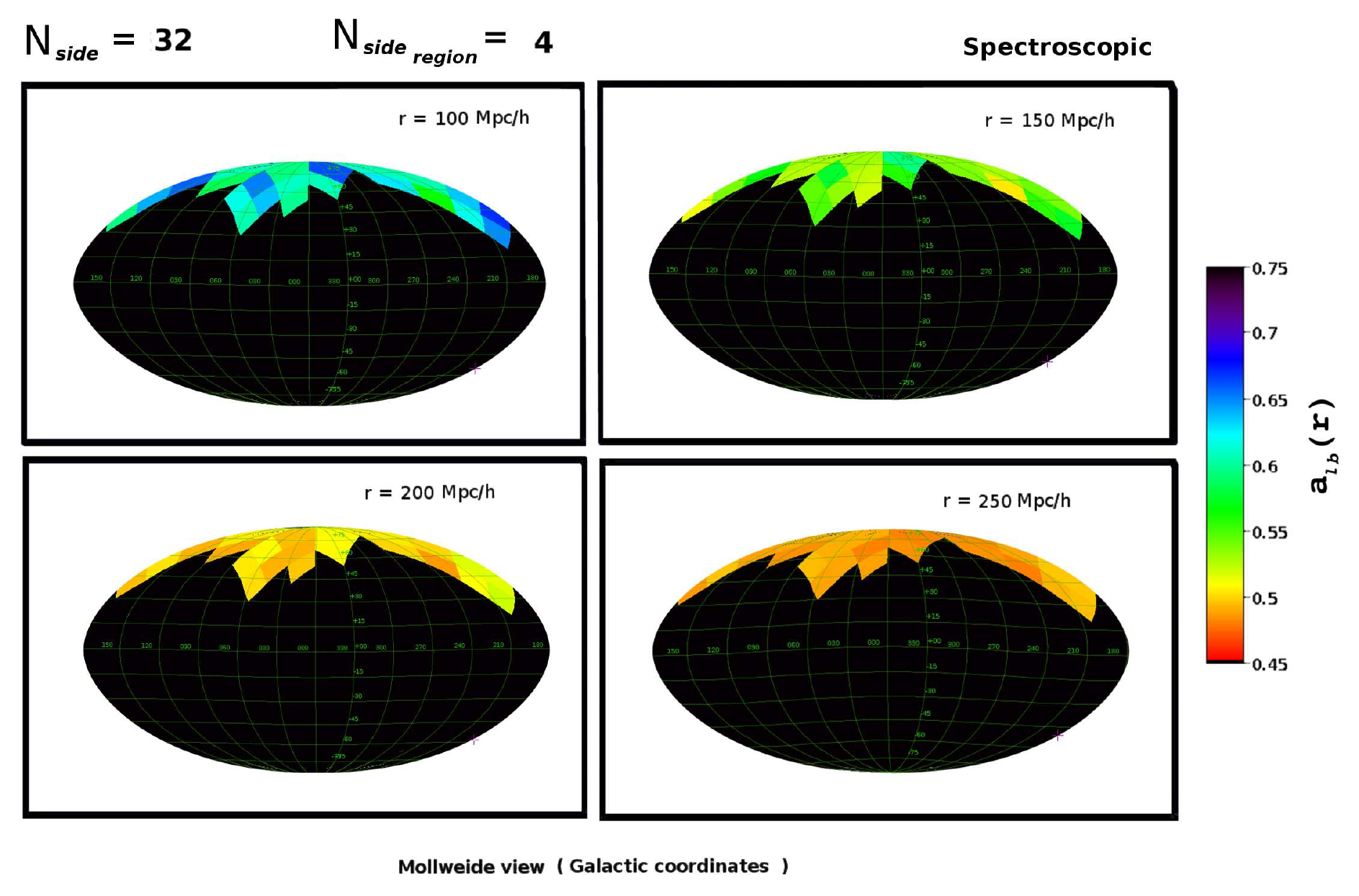}}}\\
\caption{Same as \autoref{fig:entropymap_p} but for the SDSS
  spectroscopic sky. The effective number of independent regions in
  this case is $22$.}
  \label{fig:entropymap_s}
\end{figure*}

\begin{figure*}
\resizebox{8 cm}{!}{\rotatebox{0}{\includegraphics{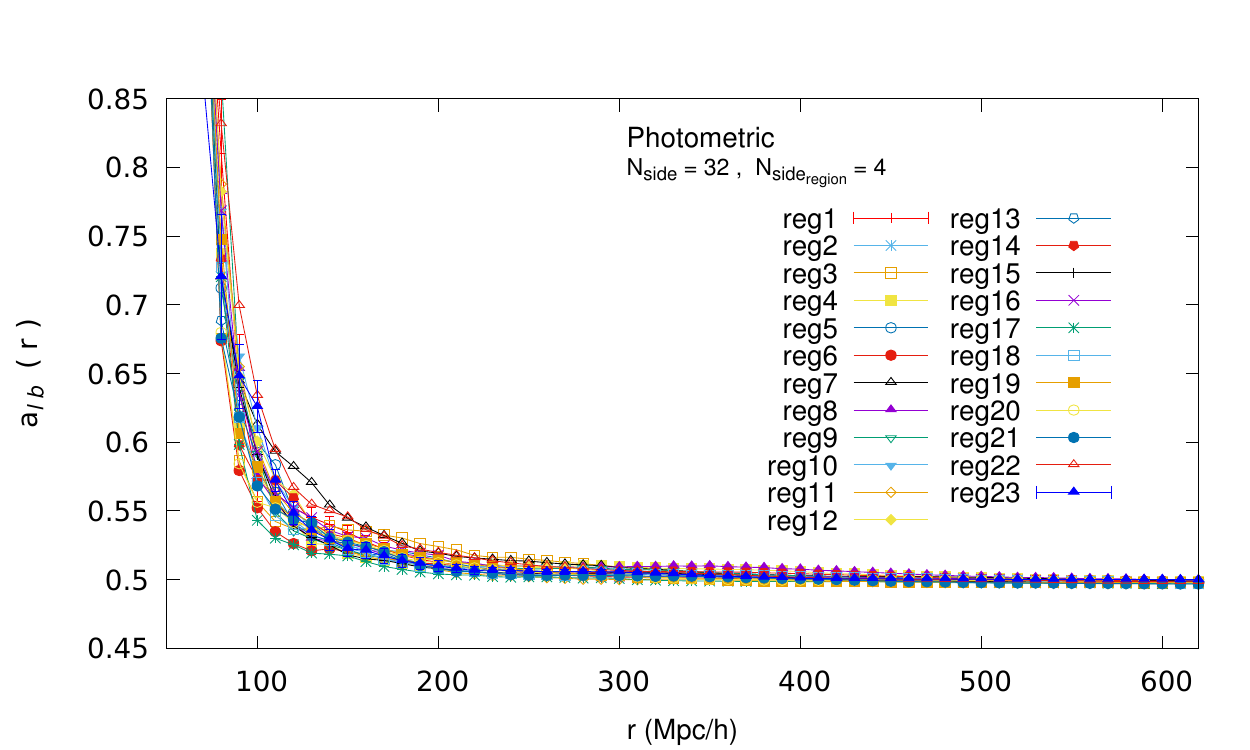}}}
\resizebox{8 cm}{!}{\rotatebox{0}{\includegraphics{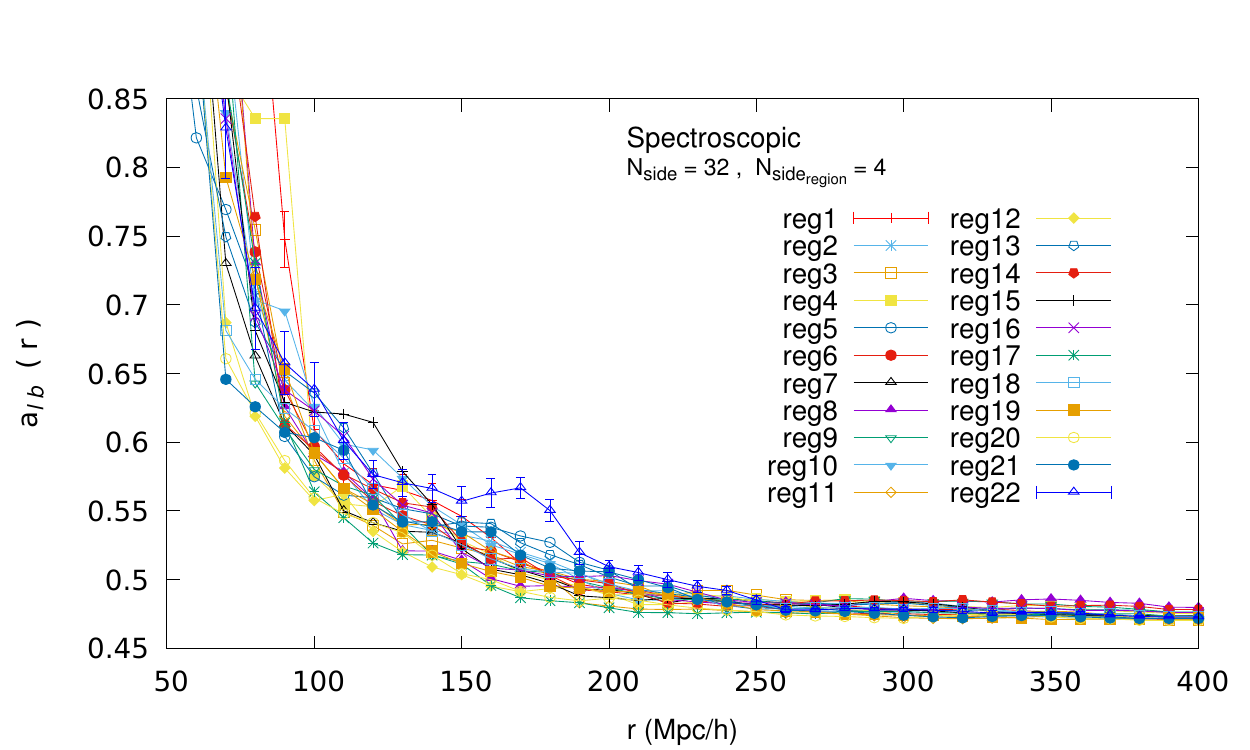}}}\\
\resizebox{8 cm}{!}{\rotatebox{0}{\includegraphics{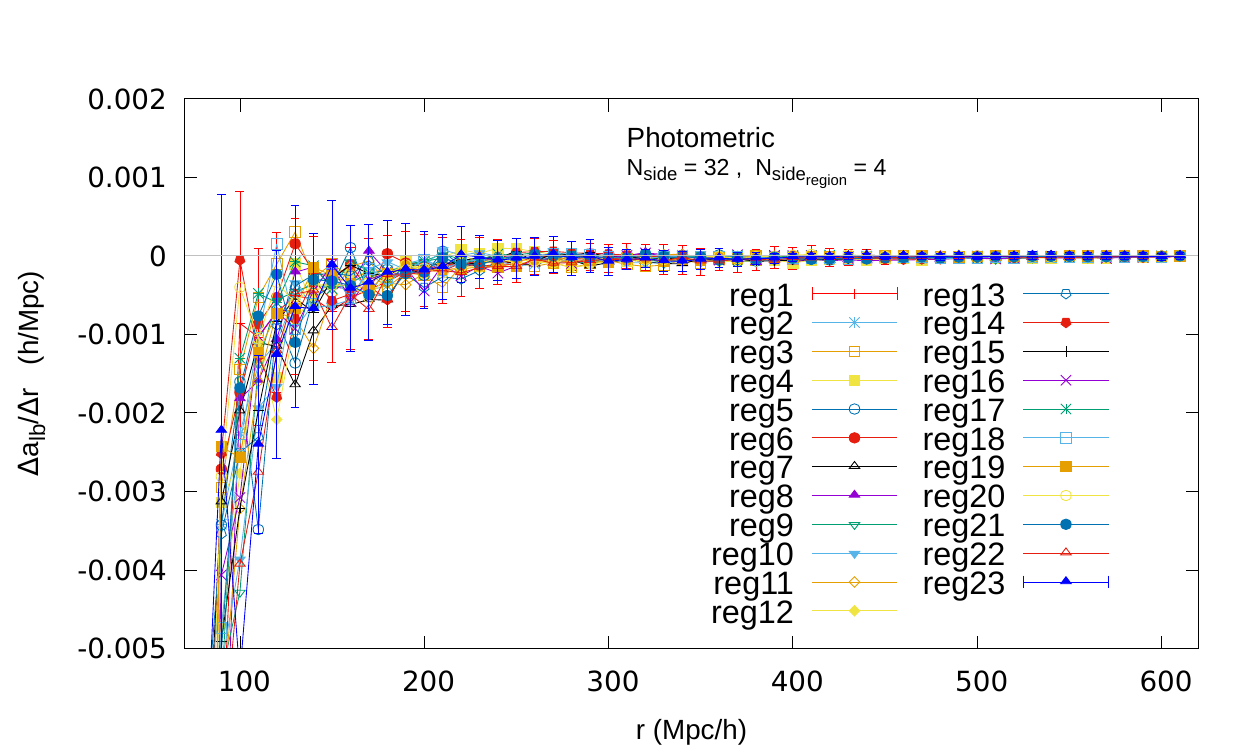}}}
\resizebox{8 cm}{!}{\rotatebox{0}{\includegraphics{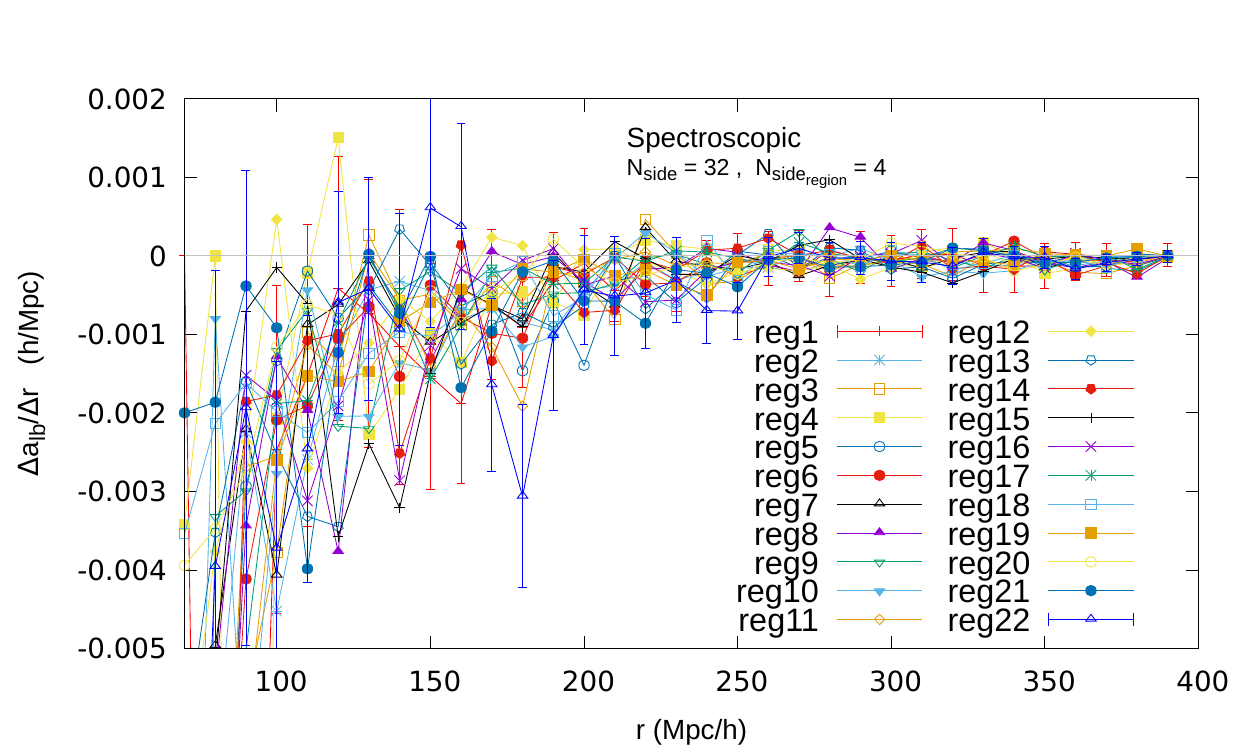}}}\\
\caption{The top left and right panels show anisotropy $a_{lb}(r)$ as
  a function of radial distance $r$ for each independent region, with
  each region corresponding to a HEALPix pixel in $N_{\rm side}=4$
  map, in the SDSS photometric and spectroscopic sky respectively. The
  two bottom panels show the rate of change of anisotropy with $r$ in
  each pixels/regions of the respective datasets. The $1-\sigma$
  errorbars in each case are estimated from the $10$ Jackknife samples
  drawn from the data.}
  \label{fig:stat_ps}
\end{figure*}

We measure the anisotropy parameter $a_{lb}(r)$ defined in
\autoref{eq:shannon2} using all the pixels inside the unmasked region
of the sky. We separately measure the anisotropy in the photometric
and the spectroscopic datasets from the SDSS at four different
resolutions with HEALPix $N_{\rm side}=4,8,16$ and $32$. The
corresponding mock galaxy catalogues from the random distributions and
the N-body simulations of the $\Lambda$CDM model are also analyzed at
the same resolution for comparison.  The results for the spectroscopic
and photometric data are shown in \autoref{fig:global_s} and
\autoref{fig:global_p} respectively.  The top left panel of
\autoref{fig:global_s} shows the variation of the anisotropy parameter
$a_{lb}(r)$ as a function of distance for the SDSS spectroscopic data
for $N_{\rm side}=4$. The results for the mock galaxy catalogues from
the random distributions and the N-body simulations of the
$\Lambda$CDM model are also shown together in this panel for
comparison with the observations. The $1-\sigma$ errorbars for the
SDSS data are estimated from the $10$ Jackknife samples drawn from the
data. The $1-\sigma$ errorbars for the mock catalogues from the random
distributions and the $\Lambda$CDM model are estimated from $10$ and
$9$ independent realizations respectively. The top right, bottom left
and bottom right panels of \autoref{fig:global_s} show the same
quantities but for $N_{\rm side}=8,16$ and $32$ respectively. In each
of these panels we see that the galaxy distribution in the SDSS is
highly anisotropic on small scales.  The observed anisotropy in the
galaxy distribution decreases with increasing length scales. It is
interesting to note that the mock galaxy catalogues from the
$\Lambda$CDM N-body simulations reproduce the observed anisotropy in
the SDSS data remarkably well for each value of $N_{\rm side}$. The
anisotropy observed in the mock catalogues from the random
distributions also decreases with increasing length scales but the
degree of anisotropy observed in these cases are noticeably smaller
than the SDSS and $\Lambda$CDM. The anisotropy observed in the random
distributions are sourced by only the Poisson noise which naturally
decreases with increasing length scales due to the increase in galaxy
counts at larger scales. On the other hand the anisotropies in the
galaxy distribution in the SDSS and the $\Lambda$CDM model are sourced
by both the Poisson noise and anisotropic gravitational
clustering. The effect of Poisson noise on the anisotropy is expected
to diminish with increasing galaxy counts in similar manner in all 3
datasets but the additional anisotropy present in the SDSS and
$\Lambda$CDM model due to the gravitational clustering would change
differently depending on the nature and strength of clustering present
in them. We find that the anisotropies in the SDSS and the
$\Lambda$CDM model decreases to a small value and plateaus out beyond
a length scale of $200 \hmpc$. Since the $\Lambda$CDM simulations
assume an isotropic background and statistically isotropic initial
perturbations, the comparison of the data with simulations tests the
validity of these assumptions. The difference between the simulations
and data in particular becomes negligible at scales $r \gtrsim 200
~{\rm Mpc/h}$.

There is excess residual anisotropy, over the isotropic Poisson
samples, present in the SDSS data and the $\Lambda$CDM
simulations. The two contributions to the anisotropy in the observed
data come from clustering in real space and from the effect of
peculiar velocities on the redshift measurements, the so called
redshift space distortions.  The SDSS galaxies are mapped in redshift
space where the peculiar velocities perturb the redshifts and distort
the galaxy distribution. The large scale coherent inflow towards the
overdense regions and outflow from the underdense regions introduce
specific anisotropic features in clustering pattern of the
galaxies. Also, the random peculiar velocities inside the virialized
bound structures elongate the structures along the line of sight
giving rise to what is popularly known as the ``Fingers of God (FOG)''
effect. Since we integrate along the radial direction, the errors in
distance measurements coming from peculiar velocities should get
averaged out. We therefore do not expect the redshift space
distortions to contribute significantly to our anisotropy parameter
$a_{lb}$. For the same reason, our results are also insensitive to the
errors in redshifts, especially in the photometric sample. We can
explicitly check this using the mock catalogues from the N-body
simulations.  We separately construct $10$ mock galaxy catalogues for
the SDSS spectroscopic sample from the $\Lambda$CDM N-body simulations
without taking into account the effect of peculiar velocities,
i.e. without the redshift space distortions. We measure the anisotropy
parameter $a_{lb}(r)$ in these samples and compare the results with
that for the mock galaxy samples prepared by taking into account the
effect of peculiar velocities, as in real observations. The results
are shown in \autoref{fig:pecvel}. We find that the anisotropy
parameter $a_{lb}(r)$ is insensitive to the redshift space
distortions. So the small residual anisotropy on large scales are not
sourced by the random peculiar velocities inside FOGs. The remaining
small difference in anisotropy between the galaxy distribution and the
Poisson distribution indicates that the galaxy distribution can not be
represented by a Poisson distribution on any length scale. We should
expect this just from linear growth. The scales of 200 Mpc entered
horizon around $z\approx 1000$ with an amplitude of $\sim 10^{-5}$ and
would have grown by a factor of $\sim 1000$ today to an amplitude of
$\sim 10^{-2}$. This is also the level of anisotropy that we see in
$a_{lb}$. We will show explicitly below that this residual anisotropy
is nothing but the linear large scale structure in the $\Lambda$CDM
universe and comparison with linear theory defines an unambiguous
scale of isotropy.

We also analyze the SDSS photometric data exactly in the same way and
present our results in \autoref{fig:global_p}. The SDSS photometric
sample covers a much larger volume and hence contains a significantly
larger number of galaxies. Interestingly, we find that the results are
very similar to those found with the SDSS spectroscopic sample. The
anisotropy in the galaxy distribution decreases with length scale in
each case and the galaxy distribution appears to be nearly isotropic
beyond a length scale of $200 \hmpc$. We note that the anisotropies
observed in the SDSS photometric sample are relatively smaller than
what is predicted by the dark matter only $\Lambda$CDM simulations on
scales below $200 \hmpc$ i.e the distribution of the galaxies in the
local Universe is more uniform than the expectations from the
$\Lambda$CDM model. But interestingly the anisotropy curves for both
the distributions flatten out on the same scales.  The differences
between the observed anisotropy in the SDSS photometric data and the
$\Lambda$CDM model may arise due to the larger uncertainties
associated with the photometric redshifts in the SDSS. We test this
possibility using the redshift estimation errors provided by
\citet{beck}. \citet{beck} use a linear fit in colour magnitude space
to describe the photometric redshift and estimate the rms error
$\delta z_{phot,i}$ in photometric redshift measurement for each
galaxies. We draw the errors on the photometric redshift for each
galaxy from Gaussians with standrd deviations $\delta z_{phot,i}$ and
simulate a set of galaxy catalogues from the primary SDSS photometric
redshift catalogue. We construct $10$ volume limited galaxy samples
from these datasets and analyze them separately. The results are
presented in \autoref{fig:errz} which indicates that the uncertainty
in the measurement of photometric redshift is unlikely to change the
anisotropy. As stated above, this is expected since we integrate along
the line of sight and any errors in redshift estimates would average
out, similar to the effect of the peculiar velocities. The difference
in dark matter only simulations and SDSS data is therefore mostly
because of the baryonic astrophysics. However, further investigations
are necessary to either support or refute this claim. One may also
note an apparent increase in the anisotropy at $\sim 300 \hmpc$ for
$N_{\rm side}=4$ and $N_{\rm side}=8$ which may arise due to the
presence of a large supercluster or void at this distance.

In the top left and right panels of \autoref{fig:global_sp}, we
compare the anisotropy in the SDSS photometric and spectroscopic
galaxy samples respectively for different $N_{\rm side}$.  We observe
a larger degree of anisotropy in the galaxy distribution at higher
resolution due to a larger contribution from the Poisson noise. We
show the rate of change of anisotropy in each case for the photometric
and spectroscopic data in the bottom left and right panels of
\autoref{fig:global_sp} respectively. We find that for both the SDSS
photometric and spectroscopic data, the observed slopes flattens out
nearly to zero on scales of $\sim 100 \hmpc$ for $N_{\rm side}=4$ and
$8$. The same trend is observed on $\sim 200 \hmpc$ for $N_{\rm
  side}=16$ and $32$ for both the datasets.

We linearly evolve the initial power spectrum upto present day and
then use it to generate particle distributions at present within the
N-body cube. These distributions represent the linear version of the
present day mass distribution. We generate $3$ such distributions and
construct $10$ independent mock samples for the spectroscopic and
photometric sample from these distributions. We analyzed them in
exactly same manner and find that a nearly constant residual
anisotropy of the order of $10^{-2}$ persists on large scales
irrespective of the chosen value of $N_{side}$. The top two panels of
\autoref{fig:slope_lin} show that there is a small offset between the
residual anisotropy observed in the SDSS and that predicted by linear
theory. The residual anisotropy in the $\Lambda$CDM mock catalogues
are consistent with the SDSS observations. 

We will use the slope of the anisotropy parameter $a_{lb}$, which is a
local quantity and insensitive to the overall normalization and
therefore cosmological parameters, to define the scale of isotropy.

In \autoref{fig:global_sp}, we see that for any $N_{side}$, the
anisotropy parameter eventually plateaus out to a small value ($\sim
10^{-2}$) on a certain scale and the corresponding slope of the
anisotropy parameter tends to zero.  However it may be noted that the
slope of the anisotropy parameter does not become exactly zero on any
length scales. In the two bottom panels of \autoref{fig:slope_lin}, we
blow up the y-axes representing the slopes of the anisotropy parameter
and find that the slopes are of the order of $10^{-4}~{\rm h/Mpc}$
beyond a length scale of $200 \hmpc$ for both the photometric and
spectroscopic sample. We calculate the slopes of the anisotropy curves
expected in linear theory and compare them to the observed values in
\autoref{fig:slope_lin} for $N_{side}=32$. We find that the slopes of
the anisotropy curves for both the photometric and spectroscopic
samples come within $1-\sigma$ errorbars of the slopes expected from
the linear theory beyond a length scale of $\sim 200 \hmpc$. We thus
conclude that the anisotropy decays in the same way as predicted by
the linear theory beyond $r\sim 200 \hmpc$. We note that the
anisotropy parameter $a_{lb}$ has contributions from all scales while
the slope is a local quantity and is only sensitive to the scale at
which it is being calculated. We therefore use the slope as the
criteria to define the scale of isotropy yielding a scale of isotropy
of $\sim 200 \hmpc$.

\subsection{THE STATISTICAL ISOTROPY IN THE PHOTOMETRIC AND SPECTROSCOPIC GALAXY SAMPLES}

Besides measuring the global isotropy across the entire contiguous
region of the SDSS photometric and spectroscopic data, we also
quantify local anisotropy in smaller regions of the sky and study how
it varies over the survey area. We divide the survey area into smaller
regions with each \emph{region} defined as a pixel of HEALPix
resolution $N_{\rm side}^{\rm region}=4$ and use resolution of $N_{\rm
  side}=32$ to study isotropy in each region individually. We obtain
$23$ and $22$ regions in the SDSS photometric and spectroscopic survey
areas respectively.  The galactic co-ordinates of these pixels are
tabulated in \autoref{tab:pixel_pos} in the Appendix.  We then compute
the degree of anisotropy in each region by calculating the entropy
associated with the counts in all $N_{\rm side}=32$ pixels inside each
region. We show the maps of $a_{lb}$ for the SDSS photometric and
spectroscopic data at different radial distances in
\autoref{fig:entropymap_p} and \autoref{fig:entropymap_s}
respectively. The radial distance provided in each panel of these
figures indicates the maximum radial distance up to which the galaxy
counts are integrated to calculate the anisotropy inside each
region. It may be noted that at a given length scale, these
anisotropies are relatively higher as compared to the global
anisotropies observed in the bottom right panels of
\autoref{fig:global_s} and \autoref{fig:global_p}. This is simply due
to the smaller sample size in each region compared to the full survey
area and therefore larger Poisson noise. We find that both the
photometric and spectroscopic SDSS sky show significant variations in
the measured anisotropy across different regions of the sky at a
radial distance of $100 \hmpc$. The variations in the local anisotropy
across regions decrease with increasing length scales and nearly cease
to exist at $\sim 200 \hmpc$.

In \autoref{fig:stat_ps}, we also show the variation in the anisotropy
parameter $a_{lb}(r)$ with distance $r$ for each of the individual 23
and 22 regions in the photometric and spectroscopic data respectively.
The observed anisotropy in each region is higher compared to the full
survey due to the larger Poisson noise in each smaller regions. The
bottom panels shows the rate of change of anisotropy in each of these
$23/22$ regions.  The results shown in \autoref{fig:stat_ps} indicate
that different parts of the SDSS photometric and spectroscopic sky
exhibit similar degree of anisotropy beyond a length scale of $\sim
200 \hmpc$ after which the anisotropy parameter $a_{lb}$ is
approximately constant. We do not find any significantly divergent
pixel beyond a length scale of $200 \hmpc$.  This suggests that the
galaxy distribution mapped by the SDSS in both the photometric and
spectroscopic redshift survey is isotropic on a length scale of $200
\hmpc$ and there are no preferred directions in the SDSS survey volume
beyond this length scale.

\section{CONCLUSIONS}
In this work, we have analyzed the SDSS photometric and spectroscopic
data with information entropy to test the isotropy of the galaxy
distribution in the local Universe. We find that the galaxy
distribution is highly anisotropic on small scales but these
anisotropies decrease with increasing radial distances. The
anisotropies predicted by the $\Lambda$CDM N-Body simulations are in
fairly good agreement with the observed anisotropies with the SDSS.
Both photometric and spectroscopic data from SDSS exhibit a small
residual anisotropy on large scales.  The $\Lambda$CDM simulations
accurately reproduce the small residual anisotropy observed on large
scales. These residual anisotropies on large scales are expected just
from the linearly evolved primordial anisotropies. To verify this, we
also study the anisotropies in the distributions generated from a
linearly evolved $\Lambda$CDM power spectrum and find a residual
anisotropy of the same order as observed in the SDSS and $\Lambda$CDM
N-body simulations. We find a small offset in the magnitude of the
residual anisotropy observed in the $\Lambda$CDM N-Body simulations
and the linear theory. We show that this offset originates from the
differences in the mass variance in the two distributions on the
corresponding length scales. To avoid the complications associated
with the offsets and normalization of the fluctuations which would be
a function of cosmological parameters, we use the slope of the
anisotropy, a local quantity, to define the scale of isotropy. We
study the slopes of the anisotropy parameter as a function of length
scales in the SDSS and linear theory and find that the slopes agree
with each other at $1-\sigma$ level beyond a length scale of $200
\hmpc$ indicating the onset of isotropy. To be precise, the decay in
anisotropy parameter $a_{lb}$ is consistent beyond this scale with the
linear perturbation theory expectation.

It may be noted that the degree of residual anisotropy depends on the
value of $N_{side}$ (\autoref{fig:global_sp}) which decides the size
of the angular patches or the volumes subtended by them at the
observer. The value of $N_{side}$ controls the magnitude of the shot
noise and hence the anisotropy.  We find that irrespective of the
choice of $N_{side}$, the galaxy distribution shows a transition to
isotropy on large scales. We would like to clarify here that the
\emph{length scale} throughout our analysis refers to the farthest
distance along the radial direction up to which the number counts are
integrated.

An analysis of the Luminous Red Galaxies (LRG) from the SDSS DR7 by
\citet{marinoni} find the scale of isotropy to be $150 \hmpc$ which is
somewhat different than the value obtained in this work. The LRG
sample analyzed by \citet{marinoni} span a redshift range $0.22<z<5$
whereas the SDSS galaxy samples analyzed here are limited to redshift
$z<0.21$. So the difference in the result arises both due to different
magnitude limits of the samples and different methods of analysis.

We test for the statistical isotropy by separately measuring
anisotropy in different sub-samples/regions of the SDSS survey and
find no evidence for a preferred direction beyond the scale $200
\hmpc$.

Our analysis indicates that the galaxy distribution shows a transition
to isotropy on a scale $\sim 200 \hmpc$ for both the SDSS photometric
and spectroscopic data. We conclude from the present analysis that the
galaxy distribution in the local Universe is indeed isotropic on a
scale of $200 \hmpc$ which reaffirms the validity of the assumption of
isotropy of the Universe on large scales.

\section{ACKNOWLEDGEMENT}
The authors thank an anonymous reviewer for useful comments and
suggestions. The authors would like to thank the SDSS team for making
the data public. The authors would like to thank Maciej Bilicki for
providing the SDSS photometric catalogue. BP would also like to thank
Maciej Bilicki for helpful comments and suggestions on the draft. SS
would like to thank UGC, Government of India for providing financial
support through a Rajiv Gandhi National Fellowship.  B.P. would like
to acknowledge financial support from the SERB, DST, Government of
India through the project EMR/2015/001037. B.P. would also like to
acknowledge IUCAA, Pune and CTS, IIT, Kharagpur for providing support
through associateship and visitors programme respectively. RK was
supported by SERB grant number ECR/2015/000078 from Science and
Engineering Research Board, Department of Science and Technology,
Govt. of India and by MPG-DST partner group between Max Planck
Institute for Astrophysics and Tata Institute of Fundamental Research,
Mumbai funded by Max Planck Gesellschaft, Germany.

\appendix
\onecolumn
\section{}
\begin{longtable}{ccc}
\caption{Galactic coordinates of the centers of the regions used for
  testing the direction dependence of the anisotropy parameter
  $a_{lb}$. These regions correspond to the HEALPix pixels at
  resolution $N_{\rm side}^{\rm region} = 4 $.}
\label{tab:pixel_pos}\\
Region & galactic longitude & galactic latitude \\
\hline
 	1  &  56.25000  & 41.81031 \\
\hline
    2  &  75.00000  & 54.34091 \\
\hline
    3  &  15.00000  & 54.34091 \\
\hline
    4  &  45.00000  & 54.34091 \\
\hline
    5  &  67.50000  & 66.44354 \\
\hline
    6  &  22.50000  & 66.44354 \\
\hline
    7  &  45.00000  & 78.28415 \\
\hline
    8  & 168.75000  & 41.81031 \\
\hline
    9  & 165.00000  & 54.34091 \\
\hline
   10  & 157.50000  & 66.44354 \\
\hline
   11  & 112.50000  & 66.44354 \\
\hline
   12  & 135.00000  & 78.28415 \\
\hline
   13  & 202.50000  & 30.00000 \\
\hline
   14  & 213.75000  & 41.81031 \\
\hline
   15  & 191.25000  & 41.81031 \\
\hline
   16  & 195.00000  & 54.34091 \\
\hline
   17  & 225.00000  & 54.34091 \\
\hline
   18  & 247.50000  & 66.44354 \\
\hline
   19  & 202.50000  & 66.44354 \\
\hline
   20  & 225.00000  & 78.28415 \\
\hline
   21  & 337.50000  & 66.44354 \\
\hline
   22  & 315.00000  & 78.28415 \\
\hline   
   23  & 180.00000  & 30.00000 \\
\hline
\end{longtable}
\bsp	
\label{lastpage}

\begin{thebibliography}{99}

\bibitem[Akrami et al.(2014)]{akrami} Akrami, Y., Fantaye, Y.,
  Shafieloo, A., et al.\ 2014, \apjl, 784, L42

\bibitem[Alam et al.(2015)]{alam} Alam, S., Albareti, F.~D., Allende
  Prieto, C., et al.\ 2015, \apjs, 219, 12

\bibitem[Alonso et al.(2015)]{alonso} Alonso, D., Salvador, A.~I.,
  S{\'a}nchez, F.~J., et al.\ 2015, \mnras, 449, 670

\bibitem[Appleby \& Shafieloo(2014)]{appleby} Appleby, S., \& Shafieloo, A.\ 2014, \jcap, 10, 070 

\bibitem[Beck et al.(2016)]{beck} Beck, R., Dobos, L., Budav{\'a}ri,
  T., Szalay, A.~S., \& Csabai, I.\ 2016, \mnras, 460, 1371

\bibitem[Bengaly et al.(2015)]{bengaly} Bengaly, C.~A.~P., Jr., 
Bernui, A., \& Alcaniz, J.~S.\ 2015, \apj, 808, 39 

\bibitem[Bengaly et al.(2017)]{bengaly17} Bengaly, C.~A.~P., Bernui, A., Ferreira, I.~S., \& Alcaniz, J.~S.\ 2017, \mnras, 466, 2799 

\bibitem[Bharadwaj et al.(2004)]{bharad04} Bharadwaj, S., Bhavsar,
  S.~P., \& Sheth, J.~V.\ 2004, \apj, 606, 25

\bibitem[Blake \& Wall(2002)]{blake} Blake, C., \& Wall, J.\ 2002,
  \nat, 416, 150

\bibitem[Briggs et al.(1996)]{briggs} Briggs, M.~S., Paciesas, W.~S.,
  Pendleton, G.~N., et al.\ 1996, \apj, 459, 40

\bibitem[Campanelli et al.(2011)]{campanelli} Campanelli, L., Cea, 
P., Fogli, G.~L., \& Marrone, A.\ 2011, \prd, 83, 103503 

\bibitem[Carlstrom et al.(2011)]{spt} Carlstrom,
  J.~E., Ade, P.~A.~R., Aird, K.~A., et al.\ 2011, \pasp, 123, 568

\bibitem[Crill et al.(2003)]{boomerang} Crill, B.~P., Ade, P.~A.~R.,
  Artusa, D.~R., et al.\ 2003, \apjs, 148, 527

\bibitem[\protect\citeauthoryear{Colles et al.}{2001}]{colles} 
Colles, M. et al.(for 2dFGRS team) 2001,\mnras,328,1039        

\bibitem[Dai et al.(2013)]{dai} Dai, L., Jeong, D., 
Kamionkowski, M., \& Chluba, J.\ 2013, \prd, 87, 123005 

\bibitem[Eriksen et al.(2007)]{eriksen} Eriksen, H.~K., Banday, A.~J.,
  G{\'o}rski, K.~M., Hansen, F.~K., \& Lilje, P.~B.\ 2007, \apjl, 660,
  L81
\bibitem[Fixsen et al.(1996)]{fixsen} Fixsen, D.~J., Cheng, E.~S.,
  Gales, J.~M., et al.\ 1996, \apj, 473, 576

\bibitem[Fowler et al.(2007)]{act} Fowler, J.~W., Niemack, M.~D.,
  Dicker, S.~R., et al.\ 2007, \ao, 46, 3444

\bibitem[Fukugita et al.(1996)]{fukugita} Fukugita, M., Ichikawa, T.,
  Gunn, J.~E., et al.\ 1996, \aj, 111, 1748

\bibitem[Gibelyou \& Huterer(2012)]{gibel} Gibelyou, C., \& Huterer,
  D.\ 2012, \mnras, 427, 1994

\bibitem[G{\'o}rski et al.(1999)]{gorski1} Gorski, K.~M.,
  Wandelt, B.~D., Hansen, F.~K., Hivon, E., \& Banday, A.~J.\ 1999,
  arXiv:astro-ph/9905275

\bibitem[G{\'o}rski et al.(2005)]{gorski2} G{\'o}rski,
  K.~M., Hivon, E., Banday, A.~J., et al.\ 2005, \apj, 622, 759

\bibitem[Gott et al.(2005)]{gott05} Gott, J.~R., III, Juri{\'c}, M.,
  Schlegel, D., et al.\ 2005, \apj, 624, 463

\bibitem[Gruppuso et al.(2013)]{grupp} Gruppuso, A., Natoli, 
P., Paci, F., et al.\ 2013, \jcap, 7, 047 

\bibitem[Gupta \& Saini(2010)]{gupta} Gupta, S., \& Saini,
  T.~D.\ 2010, \mnras, 407, 651

\bibitem[Gunn et al.(1998)]{gunn1} Gunn, J.~E., Carr, M., Rockosi, C., et al.\ 1998, \aj, 116, 3040 

\bibitem[Gunn et al.(2006)]{gunn2} Gunn, J.~E., Siegmund, W.~A., Mannery, E.~J., et al.\ 2006, \aj, 131, 2332 

\bibitem[Hanson \& Lewis(2009)]{hanlewis} Hanson, D., \&
  Lewis, A.\ 2009, \prd, 80, 063004

\bibitem[Hazra \& Shafieloo(2015)]{hazra} Hazra, D.~K., \& Shafieloo,
  A.\ 2015, \jcap, 11, 012

\bibitem[Hoftuft et al.(2009)]{hoftuft} Hoftuft, J., Eriksen, H.~K.,
  Banday, A.~J., et al.\ 2009, \apj, 699, 985

\bibitem[Jackson(2012)]{jackson} Jackson, J.~C.\ 2012, \mnras, 426,
  779
     
\bibitem[Javanmardi et al.(2015)]{javanmardi} Javanmardi, B., 
Porciani, C., Kroupa, P., \& Pflamm-Altenburg, J.\ 2015, \apj, 810, 47 

\bibitem[Javanmardi \& Kroupa(2017)]{javanmardi17} Javanmardi, B., \&
  Kroupa, P.\ 2017, \aap, 597, A120

\bibitem[Kalus et al.(2013)]{kalus} Kalus, B., Schwarz, D.~J., Seikel,
  M., \& Wiegand, A.\ 2013, \aap, 553, A56

\bibitem[Kashlinsky et al.(2008)]{kashlinsky1} Kashlinsky, A., 
Atrio-Barandela, F., Kocevski, D., \& Ebeling, H.\ 2008, \apjl, 686, L49 

\bibitem[Kashlinsky et al.(2010)]{kashlinsky2} Kashlinsky, A.,
  Atrio-Barandela, F., Ebeling, H., Edge, A., \& Kocevski, D.\ 2010,
  \apjl, 712, L81

\bibitem[Kirshner et al.(1987)]{kirshner} Kirshner, R.~P., Oemler, A.,
  Jr., Schechter, P.~L., \& Shectman, S.~A.\ 1987, \apj, 314, 493

\bibitem[\protect\citeauthoryear{{Komatsu}, {Smith}, {Dunkley}, {Bennett},
  {Gold}, {Hinshaw}, {Jarosik}, {Larson}, {Nolta}, {Page}, {Spergel},
  {Halpern}, {Hill}, {Kogut}, {Limon}, {Meyer}, {Odegard} \&
  {Tucker}}{{Komatsu} et~al.}{2011}]{wmap}
{Komatsu} E.,  {Smith} K.~M.,  {Dunkley} J.,  {Bennett} C.~L.,  {Gold} B.,
  {Hinshaw} G.,  {Jarosik} N.,  {Larson} D.,  {Nolta} M.~R.,  {Page} L.,
  {Spergel} D.~N.,  {Halpern} M.,  {Hill} R.~S.,  {Kogut} A.,  {Limon} M.,
  {Meyer} S.~S.,  {Odegard} N.,    {Tucker} G.~S.,  2011, \apjs, 192, 18

\bibitem[Land \& Magueijo(2005)]{land} Land, K., \& Magueijo, J\ 2005,
  \prl, 95, 071301

\bibitem[Lin et al.(2016)]{lin} Lin, H.-N., Wang, S., Chang, Z., \&
  Li, X.\ 2016, \mnras, 456, 1881
  
\bibitem[Marinoni et al.(2012)]{marinoni} Marinoni, C., Bel, J., \&
  Buzzi, A.\ 2012, \jcap, 10, 036

\bibitem[Meegan et al.(1992)]{meegan} Meegan, C.~A., Fishman, 
G.~J., Wilson, R.~B., et al.\ 1992, \nat, 355, 143 

\bibitem[Moss et al.(2011)]{moss} Moss, A., Scott, D., Zibin, J.~P.,
  \& Battye, R.\ 2011, \prd, 84, 023014

\bibitem[Pandey \& Bharadwaj(2005)]{pandey05} Pandey, B., \&
  Bharadwaj, S.\ 2005, \mnras, 357, 1068

\bibitem[Pandey(2013)]{pandey13} Pandey, B.\ 2013, \mnras, 430, 
3376 
\bibitem[Pandey(2016)]{pandey16a} Pandey, B.\ 2016, \mnras, 462, 1630 

\bibitem[Pandey(2016)]{pandey16b} Pandey, B.\ 2016, \mnras, 463, 4239

\bibitem[Pandey(2017)]{pandey17} Pandey, B.\ 2017, \mnras, 468, 1953

\bibitem[Penzias \& Wilson(1965)]{penzias} Penzias, A.~A., \& Wilson,
  R.~W.\ 1965, \apj, 142, 419

\bibitem[Planck Collaboration et al.(2014a)]{adeplanck1} Planck
  Collaboration, Ade, P.~A.~R., Aghanim, N., et al.\ 2014, \aap, 571,
  A23

\bibitem[Planck Collaboration et al.(2016b)]{adeplanck2}
  Planck Collaboration, Ade, P.~A.~R., Aghanim, N., et al.\ 2016,
  \aap, 594, A16

 \bibitem[Planck Collaboration et al.(2016a)]{adeplanck3} Planck
   Collaboration, Ade, P.~A.~R., Aghanim, N., et al.\ 2016, \aap, 594,
   A13

\bibitem[Schwarz et al.(2004)]{schwarz1} Schwarz, D.~J., Starkman,
  G.~D., Huterer, D., \& Copi, C.~J.\ 2004, \prl,
  93, 221301

\bibitem[Schwarz \& Weinhorst(2007)]{schwarz2} Schwarz,
  D.~J., \& Weinhorst, B.\ 2007, \aap, 474, 717
\bibitem[Shandarin \& Zeldovich(1989)]{sz1989} Shandarin, S.~F., \& Zeldovich, Y.~B.\ 1989, Reviews of Modern Physics, 61, 185 
\bibitem[Shannon(1948)]{shannon48} Shannon, C. E. \ 1948, Bell
System Technical Journal, 27, 379-423, 623-656

\bibitem[Scharf et al.(2000)]{scharf} Scharf, C.~A., Jahoda, K.,
  Treyer, M., et al.\ 2000, \apj, 544, 49

\bibitem[Smoot et al.(1992)]{smoot} Smoot, G.~F., Bennett, C.~L.,
  Kogut, A., et al.\ 1992, \apjl, 396, L1

\bibitem[Strauss et al.(2002)]{strauss} Strauss, M.~A., Weinberg,
  D.~H., Lupton, R.~H., et al.\ 2002, \aj, 124, 1810

\bibitem[Szapudi et al.(2015)]{szapudi} Szapudi, I., Kov{\'a}cs, A.,
  Granett, B.~R., et al.\ 2015, \mnras, 450, 288

\bibitem[Watkins et al.(2009)]{watkins} Watkins, R., Feldman, H.~A.,
  \& Hudson, M.~J.\ 2009, \mnras, 392, 743

\bibitem[Wilson \& Penzias(1967)]{wilson} Wilson, R.~W., \& Penzias,
  A.~A.\ 1967, Science, 156, 1100

\bibitem[Wu et al.(1999)]{wu} Wu, K.~K.~S., Lahav, O., \& Rees,
  M.~J.\ 1999, \nat, 397, 225

\bibitem[Yoon et al.(2014)]{yoon} Yoon, M., Huterer, D., Gibelyou, C.,
  Kov{\'a}cs, A., \& Szapudi, I.\ 2014, \mnras, 445, L60

\bibitem[York et al.(2000)]{york} York, D.~G., et al.\ 2000, \aj,
  120, 1579
\bibitem[Zeldovich(1970)]{zeldovich1970} Zeldovich, Y.~B.\ 1970, \aap, 5, 84 

\end{thebibliography}
\end{document}